\documentclass[conference]{IEEEtran}
\usepackage{cite}
\usepackage{amsmath,amssymb,amsfonts}
\usepackage{algorithmic}
\usepackage{graphicx}
\usepackage{textcomp}
\usepackage{url}

\usepackage{bm}

\def\BibTeX{{\rm B\kern-.05em{\sc i\kern-.025em b}\kern-.08em
    T\kern-.1667em\lower.7ex\hbox{E}\kern-.125emX}}

\begin{document}

\title{A Systematic Literature Review of Cyber Security Monitoring in Maritime}

\author{
\IEEEauthorblockN{Risto Vaarandi}
\IEEEauthorblockA{\textit{Department of Software Science} \\
\textit{Tallinn University of Technology}\\
Tallinn, Estonia \\
ORCID: 0000-0001-7781-5863}
\and
\IEEEauthorblockN{Leonidas Tsiopoulos}
\IEEEauthorblockA{\textit{Department of Software Science} \\
\textit{Tallinn University of Technology}\\
Tallinn, Estonia \\
ORCID: 0000-0002-3994-3810}
\and
\IEEEauthorblockN{\textsuperscript{} Gabor Visky}
\IEEEauthorblockA{\textit{Department of Software Science} \\
\textit{Tallinn University of Technology}\\
Tallinn, Estonia \\
ORCID: 0000-0003-0416-3985}
\and
\IEEEauthorblockN{Muaan Ur Rehman}
\IEEEauthorblockA{\textit{Department of Software Science} \\
\textit{Tallinn University of Technology}\\
Tallinn, Estonia \\
ORCID: 0009-0000-2656-0127}
\and
\IEEEauthorblockN{Hayretdin Bah\c{s}i}
\IEEEauthorblockA{\textit{Department of Software Science} \\
\textit{Tallinn University of Technology, Tallinn, Estonia \&}
\IEEEauthorblockA{\textit{School of Informatics, Computing and Cyber Systems} \\
\textit{Northern Arizona University,}
United States}}
ORCID: 0000-0001-8882-4095}

\maketitle

\begin{abstract}
In recent years, many cyber incidents have occurred in the maritime sector, targeting the information technology (IT) and operational technology (OT) infrastructure. 
One of the key approaches for handling cyber incidents is cyber security monitoring, which aims at timely detection of cyber attacks with
automated methods.
Although several literature review papers have been published in the field of maritime cyber security, none of the previous studies has focused on cyber security monitoring.
The current paper addresses this research gap and surveys the methods, algorithms, tools and architectures used for cyber security monitoring in the maritime sector. For the survey, a systematic literature review of cyber security monitoring studies is conducted
following the Preferred Reporting Items for Systematic reviews and Meta-Analyses (PRISMA) protocol. 
The first contribution of this paper is the bibliometric analysis of related literature and the identification of the main research themes in previous works. For that purpose, the paper presents a taxonomy for existing studies which highlights the main properties of maritime cyber security monitoring research. The second contribution of this paper is an in-depth analysis of previous works and the identification of research gaps and limitations in existing literature. 
The gaps and limitations include several dataset and evaluation issues and a number of understudied research topics.
Based on these findings, the paper outlines future research directions for cyber security monitoring in the maritime field.
\end{abstract}

\begin{IEEEkeywords}
cyber security monitoring in maritime, cyber security monitoring, maritime cyber security, maritime, literature review
\end{IEEEkeywords}

\section{Introduction} \label{Introduction}

Modern societies are heavily depending on shipping and port industries, and maritime services and infrastructures are often regarded as critically important \cite{Bueger2023}.
During the recent decades, maritime systems have become increasingly digitalised and interconnected \cite{Kechagias2022,Bolbot2022,BenFarah2022}.
For example, a modern vessel network relies heavily on network technologies such as Ethernet to connect the components of the ship's navigational system. As another example, the ship's Cyber-Physical System (CPS) is similarly dependent on solutions and technologies found in traditional IT systems.

Reliance on IT technologies has introduced similar cyber security issues into the maritime domain that can be found in traditional IT systems -- like organisational computer networks, maritime systems are susceptible to cyber attacks that can disrupt normal operations for longer periods, thus inflicting significant financial damage and potentially leading to catastrophic consequences. For example, in 2017, the Maersk logistics company network was attacked by NotPetya malware which brought many port terminals of the company to a halt \cite{Bronk2022}. Although the company managed to resume normal operations swiftly, it was estimated that the financial losses inflicted by the malware could have reached as high as 300 million US dollars \cite{MaerskCase}. 

Unfortunately, the above incident is not a rare event, but several similar cyber attacks have been conducted against maritime infrastructures. For example, in a recent paper Afenyo and Caesar describe 12 similar security incidents in 15 large ports \cite{Afenyo2023}, and according to their report, in 2017 large ports experienced 12 cyber attacks per day. Also, the average ransom paid due to ransomware attacks against maritime systems is 3.1 million US dollars \cite{Afenyo2023}. Additionally, cyber attacks directly affecting the control of ships or monitoring of ship traffic have been recently reported. For example, in 2016 a cyber attack misdirected two navy vessels in the Persian Gulf \cite{mnar2016iran}. In a case from 2017, cybercriminals gained access to the navigation systems of a container vessel owned by a German company, which had a capacity of 8250 TEUs \cite{containerShipHack2017}. Moreover, an Italian base station of Automatic Identification System (AIS) experienced a ship spoofing incident near Elba Island, where thousands of fake ships suddenly appeared and affected vastly the accurate monitoring of the maritime traffic in the vicinity \cite{Androjna2020}.

In response to the escalating cyber threats in the maritime sector, leading maritime organisations like the International Maritime Organisation (IMO) and the Baltic and International Maritime Council (BIMCO) have been proactive in developing guidelines and recommendations to bolster maritime cyber security. The IMO, in particular, has integrated cyber security into its safety management systems under the ISM Code. This integration, effective from January 2021, mandates ship owners and managers to assess cyber security risks and develop necessary countermeasures to address these risks as part of their safety management systems \cite{IMO-ISM}. BIMCO, the world’s largest international shipping association, has also released several guidelines on cyber security, including the ’Guidelines on Cyber Security Onboard Ships’ \cite{BIMCO}. These guidelines are designed in collaboration with other industry associations and bring forward a comprehensive framework for cyber security risks management in maritime operations. The framework covers various aspects of cyber security, from identifying threats and vulnerabilities to implementing effective risk management processes.

Despite the recent incidents and the increasing volume of recommendations and guidelines by leading maritime organisations, Kechagias et al. \cite{Kechagias2022} pointed out that cyber security remains a relatively new concept for the maritime industry, with many companies having low cyber security awareness and immature cyber security risk management culture. Furthermore, Afenyo and Caesar state in their paper \cite{Afenyo2023} that current study programs of maritime educational institutions offer insufficient cyber security knowledge to their students and trainees. These factors have led to low adoption of cyber attack detection and mitigation solutions in the maritime domain and to insufficient amount of academic research conducted in this field.

Cyber attack detection and mitigation is further complicated by unique features of maritime systems (see \cite{1_8602669, 10_9850324, 15_9478189} for a more detailed discussion). Maritime systems usually consist from a number of different IT and OT networks that connect a wide variety of maritime devices which communicate over specialised protocols such as NMEA. Therefore, the detection of attacks conducted in such networks requires detailed analysis of these maritime-specific protocols by dedicated solutions. Furthermore, ships have a limited network connectivity with the shore, which leads to the need for autonomous attack detection functionality on board. Also, space limitations and other physical constraints do not often allow to operate significant computational resources (e.g., server clusters) on board of ships, introducing the need for lightweight and resource-efficient cyber security solutions.

One of the key approaches for countering cyber attacks is \emph{cyber security monitoring} (or \emph{security monitoring} for the sake of brevity), which aims at \emph{detecting cyber attacks with automated methods in a timely fashion}, allowing to mitigate the attacks in a manual, semi-automated, or fully automated way. For example, network and host-based Intrusion Detection Systems (IDSs) are widely acknowledged security monitoring technologies for real-time surveillance of computer networks and individual hosts to detect cyber attacks. Also, most security-aware organisations operate a Security Operations Centre (SOC) for collecting data from security monitoring tools to a central Security Information and Event Management (SIEM) system where it is processed and then presented to human security analysts for further action \cite{vielberth}. In addition to existing industrial solutions, many experimental approaches like Machine Learning (ML) based algorithms have been proposed for cyber security monitoring in recent academic literature (see the domain overview paper \cite{apruzzese} for more details). 

Although security monitoring technologies have been widely adopted by traditional (i.e., non-maritime) organisations and much academic research has been conducted in this field, security monitoring remains understudied in the maritime domain. 
For example, detailed cyber security guidelines by BIMCO \cite{BIMCO} touch security monitoring only briefly, mentioning the use of network IDS and malware detection for security monitoring purposes, but failing to provide more detailed implementation recommendations to build maritime SOCs and other maritime security monitoring solutions. As another example, cyber risk management guidelines by IMO \cite{IMO-CyberRisk} do not address cyber security monitoring at all. As pointed out in \cite{1_8602669}, the generic nature of existing guidelines complicates the creation of maritime security monitoring systems, and dedicated research in this domain is needed. Also, as discussed in Section \ref{bibliometry} of this paper, maritime security monitoring is a relatively new research field. Due to the novelty of this research area, no domain overview studies have been published in maritime security monitoring that would analyse existing academic literature and identify the main research topics and open challenges.

This article fills the aforementioned research gap and specifically targets the research on security monitoring in the maritime domain. The purpose of the current study is to provide a Systematic Literature Review (SLR) of research papers published in the field. 
Note that our study focuses on peer-reviewed academic papers, excluding grey literature from consideration. A similar approach has been used in several recent SLRs that address maritime cyber security, risk, and safety \cite{Bolbot2022, mesa2024, ERBAS2024118059}. Using the same approach allows us to evaluate existing literature from an academic perspective and to identify research themes and gaps in analysed papers.

Our study begins with a discussion of selection criteria for relevant research that is followed by bibliometric analysis. We continue with creating a taxonomy for maritime security monitoring research and describe the trends and common themes in this research. Finally, our study provides a thorough analysis of existing research papers and identifies open issues, outlining future research directions for addressing these research gaps.

The remainder of this article is organised as follows. Section \ref{related_work} describes similar literature review papers, Section \ref{methodology} formulates the research questions of our study and discusses the selection criteria for research covered by our study, and Section \ref{bibliometry} contains the bibliometric analysis of the research papers selected for the study. Section \ref{findings} presents the taxonomy of relevant research and provides an in-depth analysis of selected research papers. Section \ref{discussion} discusses open issues and research gaps in the existing research literature, and Section \ref{conclusion} concludes the article.

\section{Related work} \label{related_work}

The purpose of this section is to provide an overview of similar literature review papers in the maritime domain (see Table \ref{slr-papers-table}). 
According to Table \ref{slr-papers-table}, existing papers have mainly focused on the analysis of past cyber incidents and ways for mitigating cyber threats in the future. For that purpose, the authors have proposed various measures, most notably by changing existing regulations and policies, and by improving the cyber security training and cyber awareness of personnel. Only one paper \cite{Bolbot2022} covers some security monitoring studies, but their treatment has remained very brief without a deeper analysis.
Although the papers from Table \ref{slr-papers-table} do not focus on maritime security monitoring, we will discuss these papers for the sake of the completeness of our study, and to illustrate that the treatment of maritime security monitoring has remained too brief in existing literature reviews.

\begin{table*}[thpb]
\caption{Maritime cyber security literature review papers}
\begin{center}
\begin{tabular}{|c|c|c|c|}
\hline
\textbf{Paper} & \textbf{Year} & \textbf{Topics} & \textbf{Analysis of Security Monitoring} \\
& & & \textbf{Research Literature} \\
\hline
\cite{mesa2024} & 2024 & cyber attacks against MSCN and practices for prevention and mitigation & missing \\
\hline
\cite{symes2024} & 2024 & cyber attacks on autonomous vessels and attack mitigation techniques & missing \\
\hline
\cite{ERBAS2024118059} & 2024 & cyber threat modelling and risk assessment in the field of maritime & missing \\
\hline
\cite{Yu2023} & 2023 & impact assessment for cyber threats, categorisation and mitigation of cyber risks & missing \\
\hline
\cite{Afenyo2023} & 2023 & analysis of cyber threats, regulations and policies & missing \\
\hline
\cite{Chaal2023} & 2023 & risk management, reliability and safety of autonomous ships & missing \\
\hline
\cite{Bolbot2022} & 2022 & maritime cyber security and risk management & brief treatment of some studies \\
& & & without a deeper analysis \\
\hline
\cite{BenFarah2022} & 2022 & analysis of past cyber incidents and potential cyber attacks against maritime systems & missing \\
\hline
\end{tabular}
\label{slr-papers-table}
\end{center}
\end{table*}

The SLR paper \cite{mesa2024} focused on cyber attacks in maritime supply chain networks (MSCN),
providing recommendations on the prevention and mitigation practices. The paper introduced a comprehensive taxonomy for cyber attacks which categorised them based on a number of factors, including the hacker group, attack type, affected systems, geopolitical impact, etc. According to the study, the key cyber security threats to MCSN include malware, Distributed Denial of Service (DDoS) and Denial of Service (DoS) attacks. For prevention and mitigation, the paper proposed a number of technical and policy-level measures (e.g., the use of multi-factor authentication, standardisation of cyber security practices, etc.), and practices for increasing cyber security awareness (e.g., development of training courses about attack prevention). The study briefly mentioned some security monitoring technologies without a deeper analysis.

Symes et al. \cite{symes2024} reviewed the literature focusing on the cyber attacks on autonomous vessels. The paper provided an overview on cyber attack types and described a number of well-known past security incidents in the maritime sector. The paper also discussed possible cyber attack mitigation techniques, mentioning some commonly used security monitoring techniques like network traffic monitoring without a more detailed treatment. For protecting autonomous vessels from cyber attacks, the study provided a number of recommendations, including employee training, regular updating and patching of systems, the use of strong passwords and multi-factor authentication, etc.

In \cite{Yu2023}, the papers on cyber security in the maritime domain were reviewed,
describing common trends in relevant literature. As commonly occurring research themes, the authors identified the impact assessment for maritime cyber threats, recognition of maritime cyber threats (e.g., through vulnerability management tools or through interviews with human personnel), and categorisation and mitigation of maritime cyber risks. The study also presented recommendations for risk mitigation, and discussed gaps in existing literature and avenues for further research. As a supplementary material to their paper, the authors published a list of 21 maritime cyber attacks that have received media coverage and public attention, providing a short description of each cyber attack. Although the study briefly mentioned couple of security monitoring approaches for cyber risk mitigation (e.g., the use of ML methods for AIS anomaly detection), these approaches were not elaborated further.

The overview paper \cite{Afenyo2023} focused on maritime cyber security threats,
analysing relevant literature in the field. According to the authors, existing papers do not provide enough data on past cyber security incidents and this complicates research on modelling these attacks. In addition, the authors stated that the research on maritime cyber attack data-sharing methods has received little attention so far. Furthermore, it was stated that existing studies have not focused on assessing the financial impact of cyber attacks. The authors also noted that the current educational system does not prepare maritime professionals well enough to handle cyber threats. Finally, the authors pointed out that existing governmental and international policies do not regulate maritime cyber security well enough. For example, regulations should facilitate timely reporting of cyber security incidents and cooperation between maritime companies to increase knowledge sharing on the nature and mitigation of cyber attacks.

Erbas et al. \cite{ERBAS2024118059} performed an SLR that critically examined and compared existing threat modelling and risk assessment methods in ship cyber security, developing a taxonomy for them. The authors analysed 25 scientific papers to understand the evolving landscape of cyber security practices for crewed and autonomous ships. Significant inconsistencies were observed in current approaches and key challenges were identified, underscoring the urgent need for standardised threat modelling and risk assessment frameworks that consider also the special requirements of autonomous ships. Furthermore, the authors highlighted disparities, as well as inadequacy in tool support for the reviewed threat modelling and risk assessment methods. The need for more systematic validation of expert knowledge was also deemed critical. The topic reviewed by Erbas et al. is complementary to security monitoring, as threat modelling and risk assessment can provide direct inputs for enhanced monitoring solutions. 

Bolbot et al. \cite{Bolbot2022} conducted an SLR for maritime cyber security. The review included a bibliometric analysis that explored the countries, authors, academic venues and prevalent topics of the publications within the scope. The research studies were categorised and reviewed under topics ranging from various technical topics (e.g., cyber risk management, design of countermeasures, penetration testing) to many interdisciplinary topics such as maritime law, training development, and cyber incident analysis. Cyber risk management constituted a significant portion of the reviewed studies. Although some security monitoring papers were discussed in the study, their treatment remained very brief, since the focus of the article was lying elsewhere.

The bibliometric review \cite{Chaal2023} analysed the research literature on the risk, reliability, and safety of autonomous ships. Whereas cyber security was not the primary focus of this paper, it was identified as an important part of the ship's safety. Also, the study indicated that the cyber security of autonomous vessels has started to receive more attention from researchers recently. However, the paper did not include the treatment of cyber security monitoring topics.

Another study by Ben Farah et al. \cite{BenFarah2022} reviewed cyber incidents in the maritime domain and provided a concise bibliographic analysis. The authors provided an extensive elaboration of maritime sub-systems located in vessels and ports, in addition to a review of the potential cyber attacks targeting these systems. Similarly to previous studies described in this section, the authors did not focus on cyber security monitoring issues.

Our article addresses a gap in the existing SLR literature on the cyber security of the maritime sector. Unlike previous analyses that have largely focused on the cyber threats and risks, our contribution specifically delves into maritime security monitoring systems, an area that has not been adequately covered thus far.

\section{Methodology} \label{methodology}

In this section, the systematic literature review process will be discussed, describing the undertaken methodological approach together with the inclusion and exclusion criteria for relevant research, as well as the overarching review objectives.
    
    \subsection{Aims and research questions} \label{research_questions}
        An SLR is a methodological approach for collecting studies that constitute the research literature in the given field, systematically investigating this research field, answering focused research questions, and identifying the research gaps and research directions in the field \cite{KITCHENHAM20097}. Concretely, this article provides answers to the following research questions: \\
        
        \textit{RQ1: What are the statistical characteristics of the papers that constitute the relevant literature?} Analysing the literature through bibliometric methods will provide insights on the publication timeline, the types of venues in which the works were presented, and other properties of related work. RQ1 is answered by the discussion in Section \ref{bibliometry}.
        
        \textit{RQ2: What research themes and topics are the papers focusing on, and what is the nature of research presented in the papers?} The in-depth analysis of the paper topics will be supported by the paper taxonomy created by us, which is essential for facilitating a meaningful discussion. RQ2 is answered by the analysis in Section \ref{findings}.
        
        \textit{RQ3: What are the main shortcomings and limitations of the papers?} A thorough examination of the papers will reveal potential research gaps that can be addressed in future studies. RQ3 is answered by the discussion in Section \ref{discussion}.
    
    \subsection{Survey Method}
        
        During the development of this literature review the PRISMA protocol \cite{prisma_protocol, prisma_protocol_update} was followed. 
        PRISMA is a widely used protocol for conducting an SLR and has been employed by several maritime cyber security literature review papers discussed in Section \ref{related_work} \cite{Bolbot2022, ERBAS2024118059, symes2024, mesa2024}.
        According to the PRISMA protocol, an SLR should have clearly defined objectives or research questions (see Section \ref{research_questions}), eligibility criteria for relevant literature (discussed in Section \ref{eligibility}), a search strategy for finding relevant studies and selection process (discussed Section \ref{search-strategy}), and data extraction process (discussed in Section \ref{data-extraction}).

    \subsubsection{Eligibility Criteria} \label{eligibility}
        Initially, criteria need to be set that guide the search process and help to filter out irrelevant works. In this article, papers written in English and published in peer-reviewed journals and conference/workshop proceedings were included. When searching for relevant papers, we did not set an age limit for publications (i.e., we did not exclude papers published before a specific date).
        Papers that did not focus on maritime cyber security monitoring were excluded, just like papers focusing on non-IT-related methods from the fields of electrical engineering, physics, and other fields not related to IT.
        Editorials, opinions, keynotes, abstracts, tutorial summaries, position papers, panel discussions, technical reports and posters were also excluded. 
        The duplicated publications because of different versions were eliminated from consideration, with only the most recent version being included. 
        Since this SLR considers peer-reviewed academic papers, grey literature was not covered in this research.

        Inclusion and exclusion criteria are summarised below. For a paper to be included, \emph{all} inclusion criteria described in Section \ref{Inclusion} must be met. On the other hand, a paper is excluded if it matches \emph{any} of the exclusion criteria from Section \ref{Exclusion}. 
        
        \paragraph{Inclusion Criteria} \label{Inclusion}
            \begin{itemize}
                \item Studies that are peer-reviewed journal and conference papers.
                \item Studies in the domain of cyber security in the maritime sector with a focus on security monitoring and intrusion detection.
                \item Studies that propose novel cyber security monitoring and intrusion detection algorithms and technologies, and provide details about their application in the maritime domain.
            \end{itemize}
        
        \paragraph{Exclusion Criteria} \label{Exclusion}
            \begin{itemize}
                \item Studies that do not explicitly focus on cyber security monitoring and intrusion detection.
                \item Studies without any connection to maritime.
                \item Studies that are not peer-reviewed papers.
                \item Grey literature.
                \item Studies not written in English. 
                \item Studies with their full texts not available.
                \item Studies that deal with electrical engineering, physics, and other fields not related to IT.
            \end{itemize}

        It should be pointed out that due to aforementioned criteria, a number of seemingly relevant maritime cyber security papers were excluded, because they did not focus on cyber security monitoring by contributing novel algorithms and technologies in this area. To provide a few examples of excluded papers, \cite{visky2022, ViskyTestbed2024} described maritime testbeds in the context of cyber attack tools and vulnerability testing without proposing new security monitoring approaches for detecting cyber attacks. In \cite{nganga2024sensors}, a survey was conducted among maritime SOC experts to identify challenges in maritime SOCs without proposing novel security monitoring methods for addressing these challenges. In other words, the inclusion and exclusion criteria set the focus of the current study specifically to maritime cyber security monitoring papers with original algorithmical and technological contributions.
        
    \subsubsection{Search Strategy} \label{search-strategy}
        The studies were searched according to the following steps.   
    
            \paragraph{Search for Related Publications} An initial search was applied to determine the current maritime SLR papers and to identify other significantly related publications. The most notable papers that we found have been described in Section \ref{related_work}.

        \paragraph{Keyword Selection}
            The selection of keywords heavily influences the results of the search. During the search phase, several keywords and their combinations were tried on search engines for electronic libraries. 
            All the search engines that we used supported the creation of search queries with Boolean operators (AND, OR and NOT), which helped to narrow the focus of the search and limit the number of hits.
            Also, the engines allowed to apply the search expression on the title, abstract, keywords, metadata, etc. of the papers. The time frame of the search was also definable, and the wildcard characters (such as asterisk (*) and question mark (?)) allowed further flexibility during the search.

            In the first phase of keyword selection, the keywords related to the maritime sector, like ``marine'', ``ship'', and ``navy'' were selected to define the sector. These keywords combined with the OR-operator led to 
            too many generic papers about marine systems.
            Then, the focus was narrowed by cyber security monitoring-specific keywords and phrases, like ``monitoring'', ``intrusion detection'', ``IDS'', ``attack detection'', etc. The queries which combined these keywords yielded many publications, but the majority
            of them remained out of scope.
            By adding ``cyber'' to these keywords, the focus was narrowed down further.

        \paragraph{Automated Search on Bibliographical Sources}
            During the initial stage of the automated search on bibliographical sources several trials were applied, using different combinations of keywords according to the main objectives of the review, with the aim to construct an optimal keyword set. The final keyword set was then used to construct the search query to be executed in various search engines of electronic libraries. 
            The following electronic libraries were used to identify the publications relevant for the study: ACM Digital Library, IEEE Xplore, Elsevier Scopus, and Web of Science.
            Relevant publications were searched with the following query from all libraries: \textit{cyber AND (monitoring OR ``intrusion detection'' OR ``anomaly detection'' OR ``attack detection'' OR ids) AND (maritime OR navy OR ship OR marine OR sea OR ais)}

\begin{figure*}[htbp] % OK, updated
                \begin{center}
                    \includegraphics[width=0.9\linewidth]{"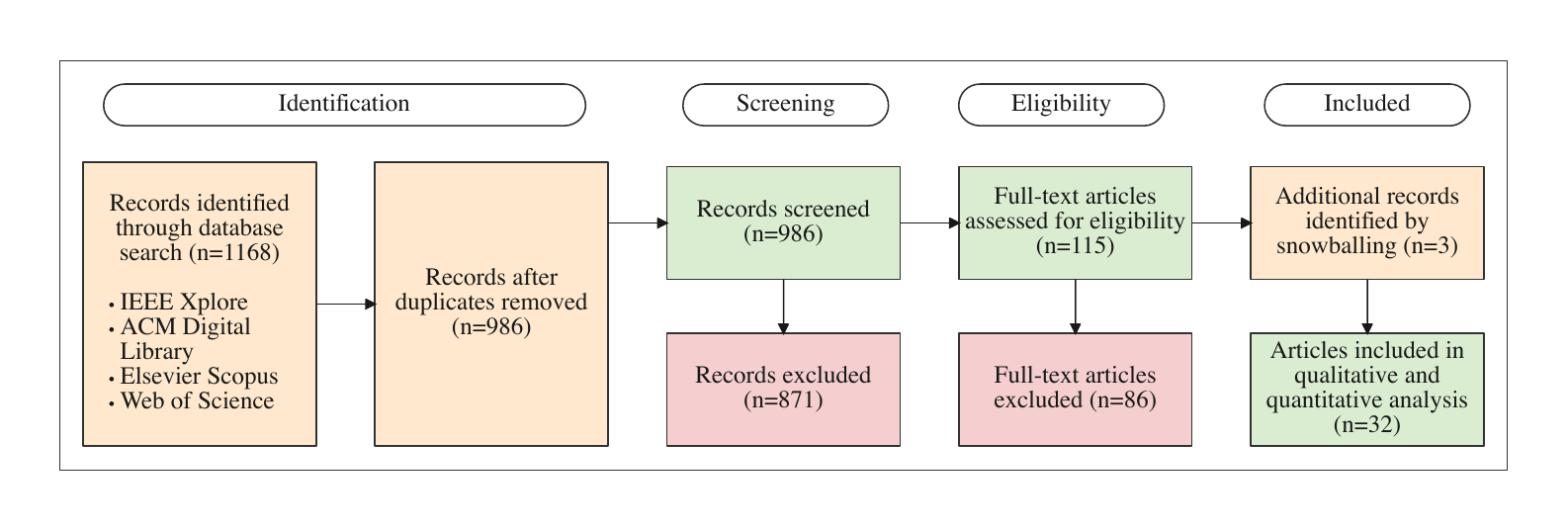"}
                    \vspace*{-3mm}
                    \caption{The use of PRISMA protocol for identifying relevant publications} 
                    \label{fig:sampling}
                \end{center}
            \end{figure*}

        \paragraph{Paper Selection Process}
            The paper selection process has been summarised in Figure \ref{fig:sampling}.
            The automated search from all libraries resulted in 1168 papers. After the library search, the four co-authors processed one library each in the form of independent work, and the findings were added to a shared database that helped to indicate the duplicated items. After the removal of the duplicates, 986 papers remained for further analysis. 

            The next step in the selection of relevant papers was the analysis of the title and the abstract. If a paper was irrelevant based on the title and/or the abstract, it was excluded (In case of doubt, the researcher read the full text). If the title and abstract were in the scope, the paper became a candidate. This part of the selection resulted in 115 candidates.
            
            If a paper remained relevant, its main properties were noted to support our findings. 
            After the full-text reading of the candidates, we had a set of papers conforming to the inclusion criteria. 
            The procedure ended in 29 papers and snowballing identified 3 additional studies, yielding 32 publications for a detailed analysis.
            At this point, the bibliometric analysis was written, and the methodology for analysis was developed.

    \subsubsection{Data Extraction and Analysis} \label{data-extraction}
                
        Following the SLR protocol, at this stage, the papers were fully assessed and relevant data were extracted relating to:
        
        \begin{itemize}
            \item Publication venue (e.g., name of the journal).
            \item Publication time and detailed author information.
            \item Research topic of the publication.
            \item The nature of the proposed security monitoring method.
            \item The nature of the maritime environment the proposed method was designed for.
            \item The nature of the data processed by the proposed method.
            \item The nature of the cyber attacks and anomalies detected by the proposed method.
            \item Performance evaluation details for the proposed method.
            \item The nature of the performance evaluation dataset and its availability information.
            \item Computational cost analysis for the proposed method.
            \item Availability information for the implementation of the proposed method. 
        \end{itemize}

\section{Bibliometric analysis} \label{bibliometry}
    
    To answer RQ1 (see Section \ref{research_questions}),
    this section presents the bibliometric analysis of relevant research.

\begin{figure}[htbp]
        \begin{center}
            \includegraphics[width=0.95\linewidth]{"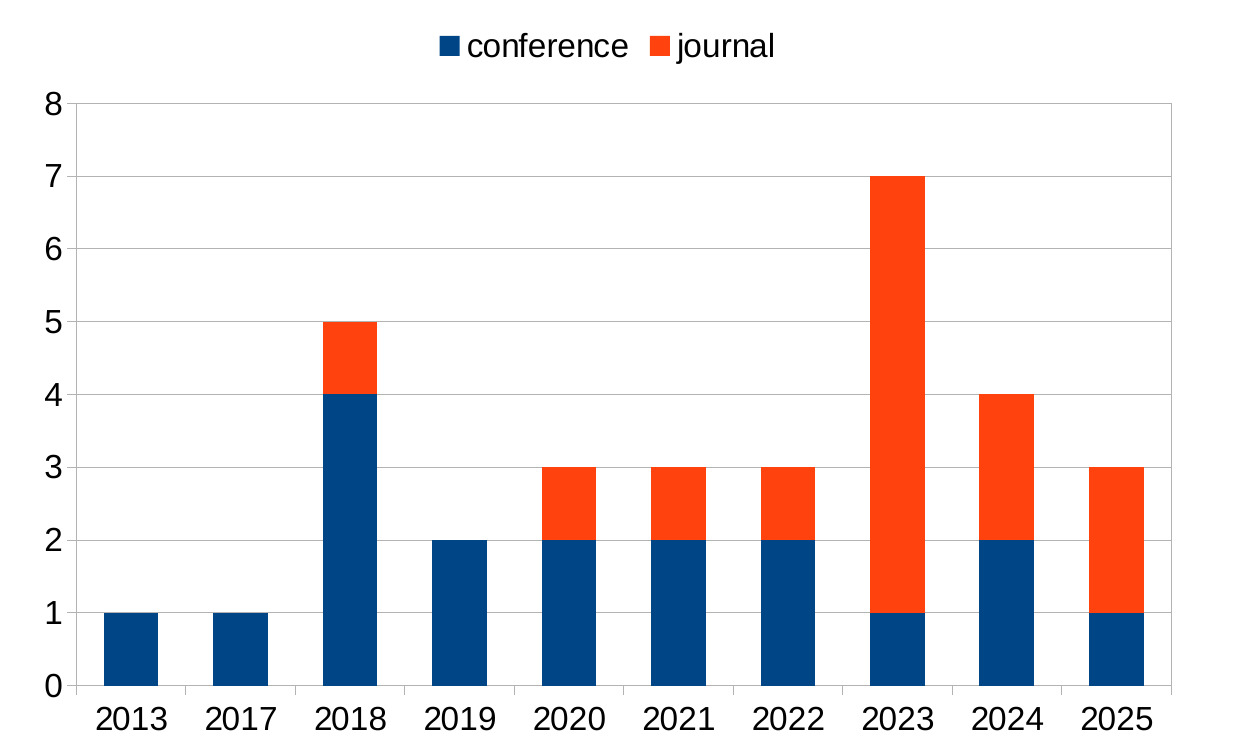"}    
            \caption{Publications per year} 
            \label{fig:years}
        \end{center}
    \end{figure}

    In  Figure \ref{fig:years}, the annual distribution of the identified papers is shown, which does not reveal a clear pattern. 
    However, an interesting observation can be made: the number of publications started rising after the major cyber incidents in 2017 within the maritime sector (e.g., the cyber attack against Maersk described in Section \ref{Introduction}). 
    As can be seen from Figure \ref{fig:years}, 30 research papers out of the 32 originate from 2018--2025, whereas only two research papers have been published before 2018. In other words, \emph{maritime security monitoring is a relatively new research domain} which has started to receive more attention during the last 6--7 years.

    According to Table \ref{table:types}, conference papers form a slight majority of the literature. On the other hand, journal articles are dominant among recent papers from 2023--2025. This indicates the increasing quality of relevant research.
    The most prominent publisher is IEEE with 24 papers (i.e., 75\% of the analysed literature). As for conference papers, the vast majority of them (17 out of 18) have been published by IEEE, whereas only one paper originates from an ACM conference. When investigating the conference venues more closely, we found that three papers 
    were published at the \emph{International Conference on Cyber Situational Awareness, Data Analytics and Assessment}, whereas each remaining conference accounted for one paper. As for journal papers, three papers 
    appeared in \emph{IEEE Transactions on Intelligent Transportation Systems} and two papers 
    in \emph{IEEE Access}. 
    Remaining journals published one paper each.

    \begin{table}[htbp]
    \centering
        \caption{Publications per publisher} 
        \begin{tabular}{|c|c|c|}
            \hline
                \textbf{Publisher} & \textbf{Journal} & \textbf{Conference} \\
                & \textbf{article} & \textbf{paper} \\
                \hline
                IEEE  &   7  &  17  \\
                ACM &  -  & 1 \\
                Elsevier &  3  &  - \\
                MDPI &  3  &  -  \\
                Springer & 1 & - \\
                \hline \hline
                Total   &   14  &  18 \\
                \hline
        \end{tabular} 
        \label{table:types}
    \end{table}

    The 32 research papers were authored by 126 researchers, and we also tried to identify more prolific authors in the field. For that task, we used a frequent itemset mining algorithm to identify all authors who have published at least two papers. Since the purpose of the frequent itemset mining algorithm is to identify associations, that allowed us to detect not just single authors, but also prolific co-author groups.
    According to our findings, no author has published more than three papers in this research domain, and a small minority of authors (16 out of 126 or 12.7\%) have published more than one paper. That indicates that for the majority of authors, maritime security monitoring has not been a persistent focus area with a larger number of peer-reviewed academic publications. In a recent SLR paper by Bolbot et al. \cite{Bolbot2022}, a similar phenomenon has been reported -- only 17\% of authors have published more than one paper in the field of maritime cyber security. Also, we identified one co-author group with three publications, with the co-authors coming from the same country. This illustrates the fact that \emph{maritime security monitoring has received limited attention so far}, and well-established prolific research groups specifically focusing on this particular field have not yet emerged.

\section{Analysis}   \label{findings}

\subsection{Taxonomy} \label{taxonomy}

\begin{figure*}[htbp]
        \begin{center}
            \includegraphics[width=\linewidth]{"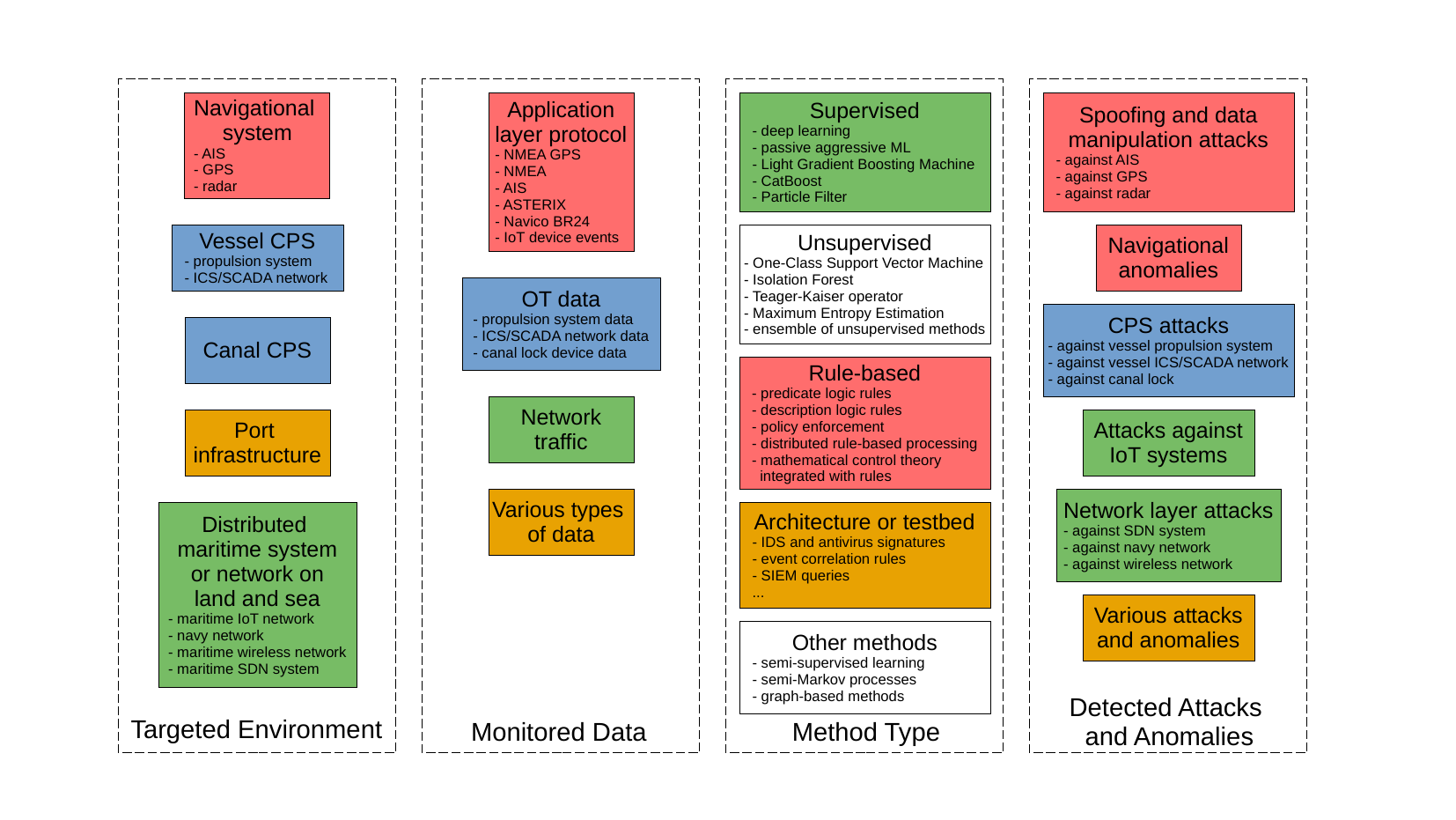"}    
            \vspace*{-10mm}
            \caption{Taxonomy for existing research literature} 
            \label{fig:taxonomy-graph}
        \end{center}
    \end{figure*}

To answer RQ2 (see Section \ref{research_questions}), we have created a taxonomy for existing research literature that is presented in Figure \ref{fig:taxonomy-graph}. Table \ref{selected-papers-table} provides detailed information about the selected 32 research papers and how they map to different taxonomy categories. As Figure \ref{fig:taxonomy-graph} illustrates, the research papers can be categorised by the environment they target, the type of data the proposed methods analyse, the type of the proposed methods, and the type of cyber attacks and anomalies that the proposed methods are able to detect. Note that these four categorisations from Figure \ref{fig:taxonomy-graph} correspond to the ``Targeted Environment'', ``Monitored Data'', ``Method Type'', and ``Detected Attacks and Anomalies'' columns in Table \ref{selected-papers-table}.

\begin{table*}[htbp]
\caption{Overview of research papers}
\begin{center}
\begin{tabular}{|c|c|c|c|c|c|c|c|}
\hline
\textbf{Paper} & \textbf{Year} & \textbf{Paper} & \textbf{Targeted} & \textbf{Monitored Data} & \textbf{Method} & \textbf{Monitoring Method} & \textbf{Detected} \\
& & \textbf{Type} & \textbf{Environment} & & \textbf{Type} & & \textbf{Attacks and} \\
& & & & & & & \textbf{Anomalies} \\
\hline
\cite{48_HossainRadar2025} & 2025 & M & navigational system & application layer protocol & supervised & deep learning & spoofing and \\
& & & & (Navico BR24 and NMEA) & & & data manipulation \\
\hline
\cite{45_SinghPopliNet2025} & 2025 & M & distributed maritime & network traffic & supervised & deep learning & IoT attacks \\
& & & system/network & & & & \\
\hline
\cite{44_AbbadNet2025} & 2025 & M & distributed maritime & application layer protocol & other & stochastic & IoT attacks \\
& & & system/network & (NMEA) & & semi-Markov processes & \\
\hline
\cite{42_HossainRadar2024} & 2024 & M & navigational system & application layer protocol & other & semi-supervised & spoofing and \\
& & & & (Navico BR24 and NMEA) & & learning & data manipulation \\
\hline
\cite{47_AgnewNet2024} & 2024 & M & distributed maritime & network traffic & supervised & CatBoost & network \\
& & & system/network & & & & layer attacks \\
\hline
\cite{41_AlgarniNet2024} & 2024 & M & distributed maritime & network traffic & supervised, & deep learning, & network \\
& & & system/network & & unsupervised & Isolation Forest & layer attacks \\
\hline
\cite{43_BaselsTestbed4Radar2024} & 2024 & T & navigational system & application layer protocol & testbed & various (different & spoofing and \\
& & & & (Navico BR24 and NMEA) & & types of IDS) & data manipulation \\
\hline
\cite{46_LongoRadar2023} & 2023 & M & navigational system & application layer protocol & rule-based & policy enforcement & spoofing and \\
& & & & (ASTERIX) & & & data manipulation \\
\hline
\cite{49_LongoTestbed2023} & 2023 & T & distributed maritime & various (network traffic, & testbed & various (IDS signatures, & various \\
& & & system/network & NMEA, etc.) & & SIEM queries, etc.) & \\
\hline
\cite{6_9796125} & 2023 & M & distributed maritime & network traffic & supervised & deep learning & IoT attacks \\
& & & system/network & & & & \\
\hline
\cite{11_9740512} & 2023 & M & distributed maritime & network traffic & supervised & passive aggressive & IoT attacks \\
& & & system/network & & & machine learning & \\
\hline
\cite{20_jmse11050928} & 2023 & M & navigational system & application layer protocol & unsupervised & ensemble of & spoofing and \\
& & & & (NMEA GPS) & & unsupervised methods & data manipulation \\
\hline
\cite{21_9617134} & 2023 & M & distributed maritime & network traffic & supervised & deep learning & IoT attacks \\
& & & system/network & & & & \\
\hline
\cite{34_Zainudin2023} & 2023 & M & distributed maritime & network traffic & supervised & deep learning & network \\
& & & system/network & & & & layer attacks \\
\hline
\cite{10_9850324} & 2022 & T & distributed maritime & various (network traffic, & testbed & various (IDS signatures, & various \\
& & & system/network & NMEA, etc.) & & SIEM queries, etc.) & \\
\hline
\cite{14_9799429} & 2022 & T & vessel CPS & various (network traffic, & testbed & various (IDS & CPS attacks \\
& & & & fieldbus traffic) & & signatures, etc.) & \\
\hline
\cite{28_info13030104} & 2022 & M & navigational system & application layer protocol & rule-based & predicate logic rules & navigational \\
& & & & (NMEA) & & & anomalies \\
\hline
\cite{15_9478189} & 2021 & M & navigational system & application layer protocol & unsupervised & One-Class Support & spoofing and \\
& & & & (NMEA GPS) & & Vector Machine & data manipulation \\
\hline
\cite{16_9728218} & 2021 & M & vessel CPS & OT (ship propulsion system) & supervised & Particle Filter & CPS attacks \\
\hline
\cite{24_WOS:000730063400003} & 2021 & M & distributed maritime & application layer protocol & supervised & Light Gradient & IoT attacks \\
& & & system/network & (IoT device events) & & Boosting Machine & \\
\hline
\cite{4_9139656} & 2020 & M & vessel CPS & OT (ship propulsion system) & unsupervised & Teager-Kaiser operator & CPS attacks \\
& & & & & & for time series analysis & \\
\hline
\cite{18_9139634} & 2020 & M & vessel CPS & OT (ship propulsion system) & other & graph-based method & CPS attacks \\
\hline
\cite{30_IPHAR2020102337} & 2020 & M & navigational system & application layer protocol & rule-based & description logic rules & navigational \\
& & & & (AIS) & & & anomalies \\
\hline
\cite{8_9065023} & 2019 & M & canal CPS & OT (canal lock device) & supervised & deep learning & CPS attacks \\
\hline
\cite{9_9032900} & 2019 & A & port infrastructure & various (network traffic, & architecture & various (IDS and & various \\
& & & & event logs, etc.) & & antivirus signatures, event & \\
& & & & & & correlation rules, etc.) & \\
\hline
\cite{1_8602669} & 2018 & A & distributed maritime & various (network traffic, & architecture & various (IDS signatures, & various \\
& & & system/network & event logs, etc.) & & event correlation rules, etc.) & \\
\hline
\cite{3_8500147} & 2018 & A & distributed maritime & various (network traffic, & architecture & various (agents with & various \\
& & & system/network & OT systems, etc.) & & different functionality) & \\
\hline
\cite{12_8512006} & 2018 & M & navigational system & application layer protocol & rule-based & distributed rule- & spoofing and \\
& & & & (AIS) & & based processing & data manipulation \\
\hline
\cite{25_10.1145/3290420.3290466} & 2018 & A & distributed maritime & various (ship navigational & architecture, & Isolation Forest & various \\
& & & system/network & data, environment data, etc.) & unsupervised & & \\
\hline
\cite{33_en11123403} & 2018 & M & vessel CPS & OT (ship SCADA network) & rule-based & mathematical control theory & CPS attacks \\
& & & & & & integrated with rules & \\
\hline
\cite{2_7956582} & 2017 & T & vessel CPS & OT (ship ICS and & testbed & dataset generation platform & - \\
& & & & SCADA network) & & without any specific method & \\
\hline
\cite{13_6735808} & 2013 & M & distributed maritime & network traffic & unsupervised & Maximum Entropy & network \\
& & & system/network & & & Estimation & layer attacks \\
\hline
\end{tabular}
\label{selected-papers-table}
\end{center}
\footnotesize{
Paper Type: M -- proposes a monitoring method; A -- proposes an architecture; T -- proposes a testbed \\  
}
\end{table*}

When using Table \ref{selected-papers-table} for analysing the mapping of papers to taxonomy categories in Figure \ref{fig:taxonomy-graph}, we discovered four non-overlapping clusters of closely related papers. The clusters cover 30 papers out of 32, and have been highlighted with different colours (red, blue, green, and orange) in Figure \ref{fig:taxonomy-graph}. If the majority of papers in some category belong to a single cluster, the category is highlighted with the colour of the respective cluster in Figure \ref{fig:taxonomy-graph}. 
Note that the ``Unsupervised'' and ``Other methods'' categories in Figure \ref{fig:taxonomy-graph} are not highlighted with any colour, since there is no cluster which would cover the majority of papers in these categories.
The four clusters from Figure \ref{fig:taxonomy-graph} are summarised below:

\begin{itemize}

\item Studies on security monitoring for navigational systems \cite{20_jmse11050928, 28_info13030104, 15_9478189, 30_IPHAR2020102337, 12_8512006, 42_HossainRadar2024, 46_LongoRadar2023, 48_HossainRadar2025} (red cluster in Figure \ref{fig:taxonomy-graph}) -- papers which propose the analysis of application layer protocol data for detecting AIS and GPS spoofing attacks, radar image manipulation attacks, and anomalies in navigational systems. 

\item Studies on security monitoring for CPSs \cite{16_9728218, 4_9139656, 18_9139634, 8_9065023, 33_en11123403} (blue cluster in Figure \ref{fig:taxonomy-graph}) -- papers which propose the analysis of OT data for detecting attacks against vessel and canal CPSs.

\item Studies on security monitoring for distributed maritime systems and networks \cite{6_9796125, 11_9740512, 21_9617134, 34_Zainudin2023, 13_6735808, 41_AlgarniNet2024, 45_SinghPopliNet2025, 47_AgnewNet2024} (green cluster in Figure \ref{fig:taxonomy-graph}) -- papers which propose the analysis of network traffic for detecting 
attacks against maritime IoT networks, maritime Software Defined Networks (SDNs), maritime wireless networks, and navy networks.

\item Studies on security monitoring architectures and testbeds \cite{10_9850324, 14_9799429, 9_9032900, 1_8602669, 3_8500147, 2_7956582, 25_10.1145/3290420.3290466, 43_BaselsTestbed4Radar2024, 49_LongoTestbed2023} (orange cluster in Figure \ref{fig:taxonomy-graph}) -- papers which have not proposed one particular security monitoring method like the studies from previous clusters, but rather a security monitoring architecture or testbed. 

\end{itemize}

The identified clusters in Figure \ref{fig:taxonomy-graph} reveal some associations between taxonomy categories which describe the nature of existing research. First, studies on security monitoring for navigational systems (red cluster) have focused on the analysis of relevant application layer protocols, and rule-based security monitoring approaches have been usually applied in the context of navigational systems. Furthermore, studies on security monitoring for distributed maritime systems and networks (green cluster) have focused on network traffic analysis, and such studies tend to employ supervised ML methods. Also, studies on security monitoring for CPSs (blue cluster) have proposed the analysis of OT data, and there is no particular method category strongly associated with this cluster. Unlike papers from other clusters, the studies on security monitoring architectures and testbeds (orange cluster) propose a range of different methods and their simultaneous use with the ability to process different types of input data and detect different attacks and anomalies.

It should be pointed out that two papers, \cite{24_WOS:000730063400003, 44_AbbadNet2025}, which are not covered by the above clusters are fairly close to one of them. 
Although these two papers focus on the analysis of application layer data instead of network traffic, they propose security monitoring methods for maritime IoT systems, and are thus closely linked to the green cluster in Figure \ref{fig:taxonomy-graph}. In the remainder of our study, we will consider these two papers together with their closest cluster.

In order to distinguish papers that have proposed security monitoring methods from the papers that have described architectures and testbeds, we divided all papers by their type (see the ``Paper Type'' column in Table \ref{selected-papers-table}). 
Letter \emph{M} denotes the papers that discuss specific security monitoring methods, letter \emph{A} the papers which describe the architecture of the entire security monitoring system, and letter \emph{T} the papers that discuss testbeds designed for security monitoring data collection, testing, and exercises. As can be seen from Table \ref{selected-papers-table}, 23 papers have focused on specific security monitoring methods \cite{4_9139656, 6_9796125, 8_9065023, 11_9740512, 12_8512006, 13_6735808, 15_9478189, 16_9728218, 18_9139634, 20_jmse11050928, 21_9617134, 24_WOS:000730063400003, 28_info13030104, 30_IPHAR2020102337, 33_en11123403, 34_Zainudin2023, 41_AlgarniNet2024, 42_HossainRadar2024, 44_AbbadNet2025, 45_SinghPopliNet2025, 46_LongoRadar2023, 47_AgnewNet2024, 48_HossainRadar2025}, while 4 papers have described architectures \cite{1_8602669, 3_8500147, 9_9032900, 25_10.1145/3290420.3290466} and 5 papers have described testbeds \cite{2_7956582, 10_9850324, 14_9799429, 43_BaselsTestbed4Radar2024, 49_LongoTestbed2023}, respectively. Most papers that concern architectures and testbeds \cite{1_8602669, 2_7956582, 3_8500147, 9_9032900, 10_9850324, 14_9799429, 43_BaselsTestbed4Radar2024, 49_LongoTestbed2023} have not proposed any novel security monitoring algorithms for analysing specific data (e.g., network traffic), but have rather employed various well-known industrial security monitoring technologies such as signature-based network IDS, antivirus, etc., which monitor a wide range of different data types (see the ``Monitored Data'' and ``Monitoring Method'' columns in Table \ref{selected-papers-table}). 
Therefore, we have set the ``Method Type'' to \emph{architecture} or \emph{testbed} for these papers in Table \ref{selected-papers-table}, because the focus of these papers lies on the entire architecture or testbed, not on one security monitoring method. The only exception is the work by Jia et al. \cite{25_10.1145/3290420.3290466} which describes the use of an Isolation Forest-based algorithm within the proposed architecture.

As can be seen from Table \ref{selected-papers-table}, security monitoring architectures received a significant amount of attention until 2019 (all 4 relevant papers were published in 2018--2019, constituting almost half of the 9 papers published before 2020). In contrast, during the more recent years after 2019, the research focus has shifted towards security monitoring methods and testbeds, with the majority of the papers (19 out of 23) discussing security monitoring methods. This finding illustrates the fact that during the earlier research period before 2020, scholars preferred to focus on the architecture of the entire security monitoring system, studying the unique challenges that arise in maritime environments. After the architectural issues and solutions for them were identified, the attention moved to particular security monitoring methods which constitute the building blocks of security monitoring system architectures.

The ``Targeted Environment'' column in Table \ref{selected-papers-table} indicates that most papers have proposed methods, architectures, and testbeds for addressing the needs of vessels or a distributed environment which involves vessels and land-based objects. In contrast, a small fraction of works target land-based environments only \cite{8_9065023, 9_9032900}.

As for the nature of monitored data (see the ``Monitored Data'' column in Table \ref{selected-papers-table}), we have used the term \emph{network traffic} for studies which have proposed network monitoring methods without considering the application layer protocol in network packets. In the rest of the article, we will call such methods network traffic monitoring methods. Almost all the studies published until 2022 were not focusing on monitoring network traffic, with the work 
\cite{13_6735808} being the only exception. Starting from 2023, a new trend can be observed where a significant proportion of papers have described such methods \cite{6_9796125, 11_9740512, 21_9617134, 34_Zainudin2023, 41_AlgarniNet2024, 45_SinghPopliNet2025, 47_AgnewNet2024}. However, these studies have been limited to specific environments like maritime IoT networks.
Also, according to Figure \ref{fig:taxonomy-graph}, the majority of papers that have proposed the analysis of application layer data have employed it for security monitoring of navigational systems, whereas the analysis of OT data is generally used in the context of CPS.

As for the types of security monitoring methods (see the ``Method Type'' column in Table \ref{selected-papers-table}), supervised and unsupervised ML approaches are the most common, while some papers have also described rule-based and other approaches. As for supervised ML, deep learning (DL) based approaches have been the most widely employed \cite{6_9796125, 8_9065023, 21_9617134, 34_Zainudin2023, 48_HossainRadar2025, 45_SinghPopliNet2025, 41_AlgarniNet2024}, whereas no unsupervised ML method has seen more frequent use than the others.
As Figure \ref{fig:taxonomy-graph} illustrates, more than half of the papers which have proposed supervised ML approaches have applied them for maritime network security monitoring, whereas rule-based approaches tend to be applied for navigational systems.

\begin{table*}[htbp]
\caption{Overview of research papers on security monitoring methods}
\begin{center}
\begin{tabular}{|c|c|c|c|c|c|c|c|c|}
\hline
\textbf{Paper} & \textbf{Year} & \textbf{Evaluation} & \textbf{Is Data} & \textbf{Is Data} & \textbf{Is Code} & \textbf{Performance} & \textbf{Computational} & \textbf{Detected} \\
& & \textbf{Data} & \textbf{Maritime} & \textbf{Publicly} & \textbf{Publicly} & \textbf{Evaluation} & \textbf{Cost} & \textbf{Attacks and Anomalies} \\
& & & \textbf{Specific} & \textbf{Available} & \textbf{Available} & & \textbf{Assessment} & \\
\hline
\cite{48_HossainRadar2025} & 2025 & RadarPWN & yes & yes & no & yes & no & radar image manipulation attacks \\
\hline
\cite{45_SinghPopliNet2025} & 2025 & SOLIDS & yes & yes & no & yes & yes & DDoS attacks against \\
& & & & & & & & maritime underwater IoT systems \\
\hline
\cite{44_AbbadNet2025} & 2025 & internal & yes & no & no & yes & no & DDoS attacks against \\
& & & & & & & & maritime IoT systems \\
\hline
\cite{42_HossainRadar2024} & 2024 & RadarPWN & yes & yes & no & yes & no & radar image manipulation attacks \\
\hline
\cite{47_AgnewNet2024} & 2024 & internal & yes & no & no & yes & yes & network attacks against \\
& & & & & & & & navy satellite networks \\
\hline
\cite{41_AlgarniNet2024} & 2024 & WSN-DS & no & yes & no & yes & no & DoS attacks against \\
& & & & & & & & maritime wireless networks \\
\hline
\cite{46_LongoRadar2023} & 2023 & internal & yes & no & no & yes & yes & radar image manipulation attacks \\
\hline
\cite{6_9796125} & 2023 & NSL-KDD & no & yes & no & yes & no & network attacks against \\
& & & & & & & & maritime IoT systems \\
\hline
\cite{11_9740512} & 2023 & CICDDoS2019 & no & yes & no & yes & yes & DDoS attacks against \\
& & & & & & & & maritime IoT systems \\
\hline
\cite{20_jmse11050928} & 2023 & MARSIM & yes & yes & yes & yes & no & GPS spoofing \\
\hline
\cite{21_9617134} & 2023 & TON-IoT & no & yes & no & yes & no & network attacks against \\
& & & & & & & & maritime IoT systems \\
\hline
\cite{34_Zainudin2023} & 2023 & X-IIoTID & no & yes & no & yes & no & network attacks against \\
& & & & & & & & maritime SDN-based systems\\
\hline
\cite{28_info13030104} & 2022 & internal & yes & no & no & no & no & anomalies in NMEA messages \\
\hline
\cite{15_9478189} & 2021 & internal & yes & no & no & yes & yes & GPS spoofing \\
\hline
\cite{16_9728218} & 2021 & internal & yes & no & no & no & no & attacks against ship's CPSs \\
\hline
\cite{24_WOS:000730063400003} & 2021 & DS2OS & no & yes & no & yes & no & application layer attacks against \\
& & & & & & & & maritime IoT systems \\
\hline
\cite{4_9139656} & 2020 & internal & yes & no & no & no & no & attacks against ship's CPSs \\
\hline
\cite{18_9139634} & 2020 & internal & yes & no & no & yes & no & attacks against ship's CPSs \\
\hline
\cite{30_IPHAR2020102337} & 2020 & internal & yes & no & no & no & no & anomalies in AIS messages \\
\hline
\cite{8_9065023} & 2019 & internal & yes & no & no & no & no & attacks against canal lock CPS \\
\hline
\cite{12_8512006} & 2018 &  MarineTraffic & yes & no & no & yes & no & AIS spoofing \\
& & AIS & & & & & & \\
\hline
\cite{25_10.1145/3290420.3290466} & 2018 & internal & yes & no & no & no & no & anomalies in the state and trajectory \\
& & & & & & & & of unmanned surface vehicle \\
\hline
\cite{33_en11123403} & 2018 & internal & yes & no & no & no & no & anomalies in ship's CPSs \\
\hline
\cite{13_6735808} & 2013 & internal & yes & no & no & no & no & network flooding attacks against \\
& & & & & & & & navy networks \\
\hline
\end{tabular}
\label{security-monitoring-papers-table}
\end{center}
\end{table*}

Note that unlike the papers on architectures and testbeds, the studies of security monitoring methods involve thorough evaluations of particular algorithms on dedicated datasets. In order to get more insights into the relevant literature,
Table \ref{security-monitoring-papers-table} provides an overview of 24 papers that propose security monitoring methods (i.e., the papers from Table \ref{selected-papers-table} with the ``Method Type'' column containing other values than \emph{architecture} or \emph{testbed}). 
Similarly to Table \ref{selected-papers-table}, Table \ref{security-monitoring-papers-table} contains the ``Detected Attacks and Anomalies'' column, but provides more detailed information about the nature of attacks and anomalies.
As this column indicates, some studies have proposed methods for detecting specific attack types. 
For example, whereas methods from \cite{6_9796125, 21_9617134, 34_Zainudin2023} can detect network attacks of various types, algorithms from \cite{11_9740512, 13_6735808, 44_AbbadNet2025, 41_AlgarniNet2024, 45_SinghPopliNet2025} are focusing on DDoS/DoS and network flooding attacks only. As another example, \cite{28_info13030104} proposes generic anomaly detection for NMEA messages, while \cite{15_9478189, 20_jmse11050928} focus on processing NMEA messages for the detection of GPS spoofing. Other methods described in papers from Table \ref{security-monitoring-papers-table} include detection of radar image manipulation attacks \cite{42_HossainRadar2024, 46_LongoRadar2023, 48_HossainRadar2025}, AIS anomaly detection \cite{30_IPHAR2020102337}, detection of AIS spoofing \cite{12_8512006}, detection of application layer attacks against maritime IoT systems \cite{24_WOS:000730063400003}, anomaly detection in the state and trajectory of unmanned surface vehicles \cite{25_10.1145/3290420.3290466}, detection of attacks against canal lock CPS \cite{8_9065023}, and detection of attacks and anomalies in the CPSs of ships  \cite{4_9139656, 16_9728218, 18_9139634, 33_en11123403}.

Table \ref{security-monitoring-papers-table} also provides some insights into the evaluation process of the proposed security monitoring methods. In older studies which used maritime-specific data for evaluations, datasets were generally private (i.e., not publicly available). Also, whenever a public dataset was utilised, it was generally not maritime-specific. However, recent studies have started a new trend of creating and using publicly available maritime-specific datasets for evaluations, and examples of such public datasets include MARSIM \cite{20_jmse11050928}, SOLIDS \cite{45_SinghPopliNet2025}, and RadarPWN \cite{RadarPWN2022}, which have been utilised in \cite{20_jmse11050928, 45_SinghPopliNet2025, 42_HossainRadar2024, 48_HossainRadar2025}.

Table \ref{security-monitoring-papers-table} also reveals several other aspects of method evaluations -- not all studies have published widely acknowledged performance metrics like \emph{precision}, \emph{recall}, and \emph{F1-score} which provide a realistic picture of the method performance (for such studies, the ``Performance Evaluation'' column in Table \ref{security-monitoring-papers-table} contains \emph{no}). Also, many studies have not released the implementation of the method (see the ``Is Code Publicly Available'' column in Table \ref{security-monitoring-papers-table}) and have not assessed the computational cost of the method (see the ``Computational Cost Assessment'' column in Table \ref{security-monitoring-papers-table}).

The following sections provide a more detailed discussion of all papers from Table \ref{selected-papers-table}. 
For structuring the discussion, we have closely followed the aforementioned clustering of the selected research papers into four clusters based on the taxonomy presented in Figure \ref{fig:taxonomy-graph}. The papers on security monitoring architectures and testbeds (with the ``Paper Type'' column set to \emph{A} or \emph{T} in Table \ref{selected-papers-table}) are covered in Sections \ref{architecture-papers} and \ref{testbed-papers}, respectively (note that such papers correspond to the orange cluster in Figure \ref{fig:taxonomy-graph}). The discussion of the remaining papers on security monitoring methods (with the ``Paper Type'' column set to \emph{M} in Table \ref{selected-papers-table}) has been divided into three subsections by the targeted environment (represented by red, blue, and green clusters in Figure \ref{fig:taxonomy-graph}) -- the papers concerning security monitoring methods for distributed maritime systems and networks are covered in Section \ref{distributed-maritime-system-papers}, papers on security monitoring methods for navigational systems are covered in Section \ref{navigational-papers}, and papers on security monitoring methods for maritime CPSs are covered in Section \ref{cps-papers}. 

\subsection{Security monitoring methods for distributed maritime systems and networks}\label{distributed-maritime-system-papers}

    Gyamfi et al. \cite{11_9740512} developed an adaptive incremental passive-aggressive ML (AI-PAML) method for a network attack detection system (NADS) in an IoT-based Maritime Transportation System (MTS) environment. A resource-efficient multi-access edge computing (MEC) setting was proposed to execute the system at network edges. To avoid data saturation problems of online learning models and facilitate the updating of the detection system, the authors provided an enhanced approximate linear dependence and advanced data updating technique to remove the irrelevant data. 
    For efficient execution, the computation-hungry algorithms of NADS were deployed through MEC servers located on the ship or onshore. 
    The experimental results were obtained by using a PC running a layer 7 DDoS simulator named DDOSIM. This was used to execute DDoS attacks against IoT devices with the aim being to achieve real-time responses from the developed NADS. The CICDDoS2019 dataset\footnote{https://www.unb.ca/cic/datasets/ddos-2019.html} was used to train the AI-PAML model. Although this dataset is quite recent and relevant for the IoT domain, it is not directly connected to the maritime domain. The authors also benchmarked the proposed model against the ``Perceptron'' and ``Stochastic Gradient Descent'' models regarding the training and prediction runtime and its latency was found to be better than the benchmark models after all models run against the DDOSIM attack.

    In \cite{21_9617134}, a monitoring system was proposed
    that used Cyber Threat Intelligence (CTI) extracted with the DL-based method, Long Short-Term Memory (LSTM) based Variational Autoencoder. The system applied the Bi-directional Gated Recurrent Unit (Bi-GRU) method for detecting attacks addressing IoT devices. According to the authors, the proposed system featured a high performance, having a better accuracy than several other similar methods from non-maritime research literature. The main drawback of the study is the fact that no maritime-specific dataset was used for the validation of the method, but rather the TON-IoT dataset\footnote{https://research.unsw.edu.au/projects/toniot-datasets} which contains network traffic of consumer IoT devices. 
    As one of the limitations, the study mentioned that the proposed DL-based method is computationally more expensive than traditional ML algorithms.

    Liu et al. \cite{6_9796125} proposed a CNN-MLP based Intrusion Detection model for MTSs which was trained via Federated Learning (FL), also called FedBatch by the authors. In the mentioned model, a Convolutional Neural Network (CNN) was utilised for data feature extractions, while Multilayer Perception (MLP) was used for attack classification to locally detect intrusion on each vessel. The study targeted IoT-based MTSs and endeavoured to address the limitations inherent in conventional learning algorithms, which consume an amount of computing resources that can not be provided in resource-constrained IoT environments. The study aimed to protect data privacy by not sharing the training data obtained locally in vessels with the cloud. Instead, the models were induced locally in the remote devices, and their parameters were shared with the cloud. 
    The authors proposed a robust and efficient model for MTS security monitoring by training and testing their model on the generic NSL-KDD dataset.\footnote{https://www.unb.ca/cic/datasets/nsl.html} 

    In \cite{13_6735808}, a method for detecting network flooding attacks was presented which harnessed a Maximum Entropy Estimation based statistical ciphertext (CT) flow analysis mechanism.
    The authors specifically focused on protecting Quality of Service (QoS) in the CT domain against DoS attacks in the context of the Automated Digital Network System (ADNS) INC III, a critical network system of the US Navy for unclassified and secret information exchange.
    The proposed approach detected anomalies in the network by using anomaly-free packets as training data and estimated their distribution as a baseline which represented normal traffic. 
    For anomaly detection purposes, the network traffic was observed and its distribution was compared with the baseline distribution.
    Furthermore, the study provided an attack response mechanism by limiting network traffic. However, in some cases, normal traffic flow was also unnecessarily limited.  
    
    Zainudin et al.\cite{34_Zainudin2023} introduced a decentralised trust aggregation solution which used blockchain to provide a secure and trusted federated IDS infrastructure. The solution offered a federated intrusion classification framework for SDN enabled marine traffic services. The scheme utilised an interplanetary file system with a network that used blockchain for proof-of-authority (PoA). An IDS model that used FL-based collaboration was also proposed, together with the lightweight intrusion detection and classification model. The solution, leveraging CNN, surpassed other comparable implementations. The effectiveness of this model was evaluated using the X-IIoTID dataset,\footnote{https://ieee-dataport.org/documents/x-iiotid-connectivity-and-device-agnostic-intrusion-dataset-industrial-internet-things} specifically curated for IoT-related research but not being maritime-specific.

    In \cite{24_WOS:000730063400003}, a supervised ML approach was proposed
    to secure sensor networks against cyber attacks and adversarial activities. The work focused on maritime IoT-based systems, addressing the attack detection in the network layer of such infrastructures.
    The paper introduced the data collection, pre-processing, model training, method optimisation, testing and its results. According to the results, the Enhanced Light Gradient Boosting Machine (Light-GBM) technique delivered excellent results on low computational cost and network bandwidth usage, which is essential in the complex marine IoT environment, that is also limited from the communication bandwidth's viewpoint. The experiments and the IoT attacks classification were conducted on the Distributed Smart Space Orchestration System (DS2OS) dataset,\footnote{https://www.kaggle.com/datasets/francoisxa/ds2ostraffictraces} an open-source but not maritime-related dataset.

    Algarni et al. \cite{41_AlgarniNet2024} explored the use of both supervised (LSTM-based DL) and unsupervised (Isolation Forest) ML methods for detecting network layer attacks in maritime wireless networks. The methods relied on the edge computing paradigm, i.e. processing data locally on network nodes. For comparative evaluation of supervised and unsupervised methods, the authors employed the WSN-DS dataset,\footnote{https://www.kaggle.com/datasets/bassamkasasbeh1/wsnds} which is not maritime specific. According to the study, LSTM-based DL outperformed Isolation Forest, demonstrating a significantly higher F1-score when detecting network attacks. 

    Agnew et al. \cite{47_AgnewNet2024} proposed a supervised method for monitoring naval submarine satellite communication networks. For creating the model of the satellite network, the authors utilised queuing theory. The method employed the CatBoost ML algorithm for detecting zero-day cyber attacks, training the algorithm on the data obtained from queuing analysis. For evaluation, simulations were used, and according to the authors, the method yielded a high precision, recall, and F1-score metric values. 
    In addition, the study provided a computational cost analysis, indicating that the proposed approach is cost efficient and requires a minimal amount of CPU time.

    The paper
    \cite{44_AbbadNet2025} suggested the use of stochastic semi-Markov processes for monitoring NMEA messages in maritime IoT networks, so that DDoS attacks in such networks could be identified in a timely fashion. The system proposed by the authors involved data collection from IoT devices and the detection of device states which reflect the likelihood of a security incident. Also, Markov chains were used to model state transitions in the network for identifying anomalies which can indicate DDoS attacks. The proposed system was evaluated by simulating normal traffic and DDoS attacks in maritime IoT systems, and according to the authors, the system demonstrated excellent performance with a low rate of false positives and false negatives.

    Singh Popli et al. \cite{45_SinghPopliNet2025} proposed a DL-based method for monitoring underwater IoT networks which involved FL. With this distributed learning approach, a pretrained global ML model was shared with devices for further local training on local data, updating the global model with parameters received from devices. For comparative evaluation of FL with traditional centralised ML, the authors employed a non-maritime-specific CICIDS2017 dataset.\footnote{https://www.unb.ca/cic/datasets/ids-2017.html} However, for evaluating the attack detection capabilities of the proposed approach, the authors used a publicly available SOLIDS dataset\footnote{https://drive.google.com/drive/u/0/folders/1IjFET0QOOZ4foEG-qzBLzTQKag6cUdiy} developed in their lab, which is reflecting normal traffic and DDoS attacks in maritime underwater IoT networks. The study also involved the computational cost analysis for the proposed method, demonstrating its suitability for real-life networks.

    \textbf{Discussion.} According to the analysis of 10 papers in this section, the current research on security monitoring methods for maritime networks has focused on specialised networks -- IoT networks have been studied in 6 papers \cite{6_9796125, 11_9740512, 21_9617134, 24_WOS:000730063400003, 44_AbbadNet2025, 45_SinghPopliNet2025}, whereas two papers have targeted military networks \cite{13_6735808, 47_AgnewNet2024}. Moreover, one paper on monitoring SDN networks \cite{34_Zainudin2023} considered an IoT-specific dataset during evaluations, making IoT networks a dominant theme in relevant research. Therefore, the monitoring of other maritime network types than IoT has not received enough attention and constitutes an understudied research area.

    As for the network monitoring methods, supervised ML methods have been the most frequently employed (proposed in 8 papers \cite{6_9796125, 21_9617134, 34_Zainudin2023, 24_WOS:000730063400003, 11_9740512, 41_AlgarniNet2024, 45_SinghPopliNet2025, 47_AgnewNet2024}), with DL-based approaches being the most prominent of them \cite{6_9796125, 21_9617134, 34_Zainudin2023, 41_AlgarniNet2024, 45_SinghPopliNet2025}. On the other hand, as argued in a recent domain overview study \cite{apruzzese}, the superiority of DL over traditional ML methods is not yet proven in the field of network intrusion detection, and DL involves a significantly higher computational cost which might not always justify its use for network monitoring (only one study \cite{45_SinghPopliNet2025} provided the relevant cost analysis). These considerations are even more pronounced for maritime systems which are often resource constrained. Therefore, the research on lightweight ML methods suitable for maritime networks is another understudied area.

    Finally, the use of supervised ML for network security monitoring is known to suffer from the issue of \emph{concept drift} \cite{apruzzese} --  since the surrounding environment changes over time, a supervised ML model will become obsolete, requiring retraining on new labeled datasets which are expensive to create. Existing works have not addressed this issue on datasets which cover longer time frames (e.g., several months), and studying ML methods for tackling concept drift (such as active learning) is a highly relevant future research direction. To this end, only the work by Gyamfi et al. \cite{11_9740512} touched on this issue by providing a technique to keep only relevant data used for the dynamic updating of the learned model, although specific validation for this aspect of the work was not provided. Another interesting direction is further study of unsupervised ML methods, which do not need labeled data.

\subsection{Security monitoring methods for navigational systems}\label{navigational-papers}

    In \cite{12_8512006}, a distributed architecture was presented for
    detecting spoofing and falsification attacks in streams of AIS data in real time. Average speed was measured on the shortest path between positions obtained from two consecutive AIS broadcast messages. Then, this value constituted the basis of the spoof detection approach implemented in the study. 
    The architecture was of the Master-Workers type, with the Master distributing the AIS messages to Workers for analysis in an efficient manner incorporating load balancing techniques.
    The approach was validated by generating a 43,912,236 AIS messages dataset from \emph{marinetraffic.com} and performing simulations by attacking the dataset in various ways. The validation dataset has not been made public. The reduction of false positives as well as the extension to more attack types was mentioned to be part of future work.
     
    Amro et al. \cite{28_info13030104} proposed a method for analysing navigational NMEA-0183 messages that communicate the data of the various onboard sensors. The objective was to identify possible anomalies and their malicious causes. For this, relevant anomaly detection algorithms were developed. To facilitate the analysis the authors employed a maritime cyber security testing environment that they previously developed \cite{NTNU_Testbed} which included the NMEA-Manipulator tool enabling the generation of the anomalies by attacking sensor data. 
    To develop the anomaly detection method, the authors first identified NMEA message types that affected navigation together with relevant message fields and their values. Potential anomalies that can appear in NMEA messages and attack techniques which can trigger these anomalies were studied using the MITRE ATT\&CK framework.\footnote{https://attack.mitre.org/} Finally, the authors investigated different approaches for implementing attack and anomaly detection for NMEA messages. 
    According to the authors, specification-based approach was best suited for that purpose, which involved the use of rules that described the normal behaviour of the system (i.e., any violation of these rules indicated an anomaly). Also, for some specific attacks that involved changing the arrival rate of NMEA messages to the system components, the authors recommended frequency-based anomaly detection that involved checking if the message rate remained within the expected boundaries.

    Iphar et al.
    \cite{30_IPHAR2020102337} analysed the weaknesses of AIS in their study. Since these weaknesses allow falsifying and spoofing AIS messages, a rule-based system was developed for anomaly detection from AIS messages. The authors proposed over 900 rules or \emph{integrity items} for assessing the integrity of AIS messages. The first order assessment rules considered only one field from a single AIS message, while the second order rules analysed several fields from a single message. The third order rules considered fields from several AIS messages of the same type, whereas the fourth order rules analysed several AIS messages of all types. Based on the truth values of rules and external contextual information, flags were set that were further used for calculating ship-related risks and their levels.
    For evaluating the proposed method, a Python-based prototype was created which utilised a database for storing AIS data, intermediate results from computations, and contextual data. The prototype was evaluated on about 24 million real-life AIS messages collected over 6 months. In the study \cite{30_IPHAR2020102337}, the authors provided a detailed description of risk evaluation for 4 cases involving anomalous AIS messages (the study considered 13 cases in total). In all 4 cases, the prototype raised one or more flags. However, the study did not evaluate the system on a larger number of cases, including the assessment of a false positive rate of the system (reflected by the precision metric in ML literature).

    The study 
    \cite{15_9478189} focused on the detection of GPS spoofing. For that purpose, an anomaly detection approach was proposed which employed a One-Class Support Vector Machine (OC-SVM) and analysed NMEA messages that carried GPS information. The authors selected OC-SVM because it was easy to implement and did not require much computational resources, making it suitable for deployment on low-end hardware. During the experiments, the authors trained OC-SVM on data collected from a maritime simulator and also on real-life data collected on a small boat. For implementing GPS spoofing during the experiments, the authors utilised a device which generated a GPS signal carrying wrong GPS information.
    According to the experiments, the OC-SVM based anomaly detection method featured a high recall and precision. To demonstrate the lightweight nature of the method and its low computational cost, the authors implemented the method on a Raspberry Pi based monitoring device that was suitable for deployment on real-life ships. Such evaluation of the actual computational cost of the proposed method and its suitability for real-life deployment sets this study apart from many other works. 
    
    Similarly to the previous paper 
    \cite{15_9478189}, the study by Spravil et al. \cite{20_jmse11050928} focused on the detection of GPS spoofing with the analysis of relevant NMEA messages. According to the authors, the analysis of NMEA messages allowed for the development of dedicated monitoring modules which could be easily integrated into the ship's network. To demonstrate the viability of that approach, the authors publicly released the implementation of the MANA (MAritime Nmea-based Anomaly detection) framework. The framework combined unsupervised anomaly detection methods into an ensemble, and to identify suitable methods, the authors investigated various GPS spoofing methods, with a number of methods originating from other domains than maritime (e.g., aviation and mobile phone networks). As a result of the investigation, five methods were selected for the ensemble.
    To assess the performance of the MANA framework, the authors created the MARSIM dataset which contained a large number of different GPS spoofing attacks and normal GPS data. Also, the dataset was released into the public domain.\footnote{https://github.com/fkie-cad/mana} 
    According to the experiments, the ensemble of five methods featured a higher recall than each method (the ensemble regarded the anomaly as detected if any of the five methods provided a relevant indication). As one of the limitations of their approach, the authors identified the need for well-chosen thresholds for each method to avoid false positive alerts. 

    Hossain et al. \cite{42_HossainRadar2024} proposed a semi-supervised ML approach for detecting attacks against maritime radar systems which involve manipulations of the radar image (e.g., freezing or rotating the image, adding an object to the image, etc.). One of the main purposes of semi-supervised ML is to lower the data labelling workload of human experts \cite{apruzzese2}. For the experiments, the authors used the RadarPWN dataset\footnote{https://doi.org/10.5281/zenodo.6805559} which is maritime specific and publicly available (more details on the dataset can be found in \cite{RadarPWN2022}). For semi-supervised ML, the authors utilised the Random Forest classifier which was initially trained on a small dataset labelled by a human expert. The classifier was then used to predict the labels for unlabelled training data, and samples with the most certain predictions were added to training data. After that, the classifier was trained again on the labelled training dataset extended with newly added samples. The whole training process consisted of 10 such training iterations, extending the training dataset during each iteration with automatically labelled samples, which reduces the workload of human experts. According to the authors, the final classifier achieved a high precision, recall, and F1-score metric values.

    The study 
    \cite{46_LongoRadar2023} described a number of attack scenarios for manipulating the maritime radar image which involved the misuse of the ASTERIX protocol. For detecting these attacks, the authors proposed a rule-based system which utilised candidate policies, where each candidate policy described the normal operating conditions of a radar. For making candidate policies adjustable for specific environments, the policies contained variables which were automatically set by the system. Also, if some candidate policies were not applicable in the given environment, they were automatically excluded from consideration. Selected policies were used for the real-time analysis of ASTERIX messages, and alerts were raised on policy violations. According to the study, the proposed system featured a high attack detection rate with a small number of false positives. In addition, the computational cost of the system was low, making it suitable for resource-constrained maritime environments. The experiments conducted in the study were supported by the MaCySTe testbed \cite{49_LongoTestbed2023}, which will be described in Section \ref{testbed-papers}.

    To avoid data privacy issues of conventional ML and DL approaches that typically require centralising data on a single server, Hossain et al.  \cite{48_HossainRadar2025} presented an FL-based approach
    with the aim to enhance the detection and classification of cyber attacks against marine radar systems. 
    The method constructed a global model by utilising the collective learning ability of several distant clients, which involved local training at clients without sharing sensitive data. For evaluation, the RadarPWN dataset was used and partitioned into numerous subsets to represent different data sources in a collective learning environment. The authors provided a comprehensive evaluation of their approach, 
    with high values reported for each performance metric. Additionally, the authors evaluated several DL-based approaches for the attack 
    detection component of their system, choosing CNN as the best fit. The authors stated some limitations of their study, e.g., 
    the need for reliable communication between clients and the central server, which might not always be practical in maritime operations due to connectivity issues.

    \textbf{Discussion.} All 8 papers discussed in this section have focused on the analysis of application layer data (e.g., NMEA or AIS messages) in order to detect AIS spoofing, GPS spoofing, image manipulation attacks against maritime radars, and navigational anomalies. As for the used methods, rule-based approaches are the most common and have been suggested in four papers \cite{12_8512006, 28_info13030104, 30_IPHAR2020102337, 46_LongoRadar2023}, whereas remaining four papers have suggested unsupervised ML \cite{15_9478189, 20_jmse11050928}, semi-supervised ML \cite{42_HossainRadar2024}, and DL \cite{48_HossainRadar2025}. When employed for security monitoring, rule-based approaches are known to be very precise \cite{apruzzese} and are thus used by many industrial security monitoring products (such as commonly used network IDS platforms like Suricata). However, creating a rule-based system requires domain experts and is a time-consuming process. One open research area is the use of rule mining algorithms in order to speed up the rule creation process.

    Also, the current research has largely focused on rule-based and unsupervised ML-based approaches, whereas only two papers analysed in this section have suggested the use of supervised and semi-supervised ML for the detection of navigational attacks and anomalies. Therefore, further study of these approaches is another open research area, as is dealing with advanced ML issues like concept drift. Finally, as mentioned in \cite{apruzzese}, rule-based and ML-based security monitoring systems are complementary, with both of them having unique advantages. Therefore, building and evaluating such hybrid systems for navigational security monitoring is another future research direction. Inductive logic programming \cite{ILP_at_30} is a potential hybrid approach which could be applied to automate the rule learning process in a supervised ML manner. 

\subsection{Security monitoring methods for maritime CPS}\label{cps-papers}

    In \cite{18_9139634}, graph-based models were utilised
    for the detection of attack propagation in CPSs. A graph model that represented the digital and physical system layers was proposed in the paper, considering the interdependencies between CPSs by using the system variables. The model was applied in a case study that involved a small-scale system consisting of a ship with a fuel tank, valve, motor and propeller which were monitored through a PLC. One scenario of normal system usage and four attack scenarios including port scan, closing the PLC and man-in-the-middle attacks were simulated in the study. 
    According to the authors, the method did not produce a false positive alert for the normal scenario, whereas one attack scenario out of four remained undetected (i.e., precision and recall were 100\% and 75\%, respectively). 
    However, the evaluation of the method was limited to the aforementioned five scenarios without testing the method on a larger dataset. 

    Yoginath et al. \cite{8_9065023} proposed a Recurrent Neural Network (RNN) based Digital Twin (DT) model to simulate an operational CPS of a Canal Lock. Their model performed real-time analysis to detect anomalies at the earliest, specifically anomalous changes in water levels within chambers through which ships pass. The study \cite{8_9065023} used an actual PLC, sensors, valves and pumps to emulate the behaviour of the Canal Lock CPS physically. The authors identified all the possible scenarios, i.e., the ship’s up and down movements, performed by Canal Lock CPS and trained a set of RNN models using the emulation data as input. Anomaly detection was performed through a difference-based method (analysis of PLC and DT predicted values). The previously DT predicted value was compared with the current PLC read value and the deviation was captured, with the larger value being regarded as an indication of the attack. 
    Furthermore, to evaluate the RNN model, the authors performed live validation of the system. However, no specific CPS attacks were executed to test the accuracy of the DT concept for anomaly detection.

    Xing et al.
    \cite{33_en11123403} presented an anomaly detection method for the Ship Information System (SIS) based on risk data analysis. Since SIS is a typical SCADA system, the authors first elaborated on all the different operational and networking components of the SIS SCADA system and then they proceeded with presenting the data anomaly detection approach based on a cooperative state space control mathematical model, for the data propagated between the sensors, the distributed controller units and the actuators. A so-called Critical State Estimation (CSE) Algorithm was proposed based on the Industrial State Modelling Language (ISML). Detection rules were formulated as ``condition \textit{implies} action'' with the condition being a boolean formula composed by various predicates indicating values that are expected by system components. The approach was validated by simulations on the ship dynamics under heading sensor signal attacks. 

    The study 
    \cite{4_9139656} focused on the use of Teager-Kaiser operator for time series analysis to detect anomalies in a ship's propulsion system in the case of cyber attacks. During the experiments, the authors collected the following time series data -- oil consumption, fuel consumption, propeller speed, and valve opening. The following attacks were conducted for evaluating the anomaly detection methods -- port scanning with \emph{nmap} tool, DoS attack with \emph{hping} tool, and exploiting the vulnerability of the ship's propulsion PLC for stopping and starting the PLC. Port scanning and DoS attacks did not influence the propulsion system, and only the attack against the PLC had a significant impact. According to the experiments, the Teager-Kaiser operator managed to identify sudden changes in the ship's propulsion system time series data. However, the evaluation results were presented in the form of graphs only for the 5-minute time period which involved a single attack against the PLC. Therefore, it is difficult to assess what would be the detection rate for many attacks of different types against the propulsion system during a longer time frame (e.g., several days or weeks), and a more detailed analysis of the proposed method was identified as future work in the study.

    The study 
    \cite{16_9728218} proposed the use of a Particle filter-based anomaly detection method for time series data collected from a ship's CPS to detect cyber attacks. In the study, the authors focused on data collected from the ship's propulsion system -- the temperature of the cylinder exhaust, cooling water pressure, and the speed of the fuel flow. For training the anomaly detection system, the authors used data collected from a new cargo ship during its first sailing week, assuming that the probability of cyber attacks during this week is very low. For evaluating the anomaly detection algorithm, data from the second half of the first sailing year was employed. For discussing the anomaly detection rate of the algorithm, the authors presented graphs for the 5,000-minute time frame (about 3.5 days) in the study. Although the graphs displayed some fluctuations in the time series data for the propulsion system, the study did not detail whether these sudden changes were the result of a cyber attack or rather a side effect of normal navigation.

    \textbf{Discussion.} As for the research themes in five analysed papers, three papers focused on the vessel propulsion system \cite{4_9139656, 16_9728218, 18_9139634}, whereas the remaining two papers targeted vessel SCADA network \cite{33_en11123403} and canal lock CPS \cite{8_9065023}, respectively. Therefore, apart from the propulsion system, other parts of the vessel CPS have received little attention so far in the context of cyber security monitoring. Similarly, research on security monitoring for land-based maritime CPSs is scarce, and these research gaps are likely to be addressed by future works.

    As for the types of security monitoring methods (see Table \ref{selected-papers-table}), no particular method types have been dominantly employed for security monitoring of maritime CPSs, and proposed methods have included supervised ML \cite{16_9728218, 8_9065023}, unsupervised ML \cite{4_9139656}, rule-based \cite{33_en11123403}, and graph-based \cite{18_9139634} methods. However, when analysing OT data from maritime CPSs, existing methods have generally not considered discriminating cyber attack related CPS malfunctions from CPS faults not caused by malicious cyber activity. Therefore, the root cause analysis for maritime CPS faults and anomalies is an unexplored research area which deserves closer attention.

    Furthermore, process-aware attacks target CPSs to induce disruptions in physical processes \cite{Khorrami_7523254}. Temporal characteristics of the processes observed on the sensors, actuators or control algorithms are utilised for the detection of malicious actions beyond the failures \cite{Khorrami_7523254}. The reviewed papers do not address this monitoring approach comprehensively. Just focusing on the control functions in the physical space via process-aware approaches may not be enough to discriminate the usual system failures from malicious actions. Process-centric indicators should be correlated with other cyber-space indicators that reveal the prior actions of the attackers before compromising the control functions. To facilitate this research, it is necessary to generate datasets that include indicators about physical and cyber spaces. 

    Finally, in order to properly study the effects of cyber attacks on maritime CPS, a highly realistic lab environment is needed which would mimic a real-life CPS as closely as possible. Apart from a few recent works \cite{14_9799429, 49_LongoTestbed2023} that will be discussed in Section \ref{testbed-papers}, the creation of realistic maritime CPS testbeds has been understudied, and can be regarded as a promising future research direction.

\subsection{Security monitoring system architectures}\label{architecture-papers}

    In \cite{1_8602669}, a maritime SOC architecture was proposed.
    The authors first provided an overview of the specific constraints of the maritime domain which complicate building a maritime SOC (for example, the limited network connectivity of the vessels with shore). The SOC architecture described by the authors took these constraints into account, consisting of vessel-based and shore-based parts. 
    The proposed vessel-based functional blocks of the SOC were the following:
    
    \begin{itemize}
        \item \emph{Network Connection Safety} allowed to safely connect monitoring sensors to monitored systems (e.g., port mirroring for network monitoring sensors);
        
        \item \emph{Network Probe Isolation} was responsible for monitoring the network (e.g., a signature-based IDS), while being isolated from other sensors; 
        
        \item \emph{Local Preprocessor} normalised and correlated events, to lessen the amount of data that needed to be sent to shore via network link with limited bandwidth; 
        
        \item \emph{Local Engine} stored events on a ship for local alerting; 
        
        \item \emph{Ship Shore Manager} acted both as a data cache and sender of data from a ship to shore, keeping the data in cache if there were issues with the network link, and being responsible for the link bandwidth management;
        
        \item \emph{Cyber Situational Awareness Console} provided a local simplified overview of the ship's cyber state to the crew. The presence of the local onboard monitoring capability allowed for addressing scenarios where connectivity with the shore-based SOC was missing and the ship's situation could not be assessed in a central land-based monitoring centre.
    
    \end{itemize}
    
    The proposed shore-based functional blocks of the SOC were the following. First, \emph{Ship Shore Manager} and \emph{Bandwidth Manager} were responsible for receiving the security data from ships and managing network connectivity with remote parties; \emph{Central Processor} was responsible for filtering and normalising the received security data, and \emph{Data Store} was a big data solution that stored the normalised security data.
    In addition to the aforementioned blocks, the shore-based SOC had other functional blocks that can also be found in traditional SOCs. For example, human operators could use specialised tools for big data analytics, searching the collected security data, creating visualisations, etc., to achieve maritime cyber situational awareness. In addition, SOC could be linked to external threat intelligence sources and incident response platforms. Finally, collected security data could be used for creating a so-called maritime-recognised cyber picture of the status of individual ships and the entire fleet.
    In their study \cite{1_8602669}, the authors mentioned that the previously described architecture has been thoroughly tested. 
    Although no detailed information was provided about the nature of these tests and the performance of the architecture,
    the architecture was validated in a later independent study \cite{10_9850324} by building a maritime exercise environment according to this architecture (see Section \ref{testbed-papers}).

    Schauer et al. \cite{9_9032900} proposed a security monitoring system architecture for ports which would integrate security data from two distinct domains -- traditional IT systems (i.e., cyber assets) and physical assets of the port. The authors proposed to collect a wide variety of security events from IT systems, including firewall events, antivirus events from end user devices, and event logs of applications and servers. To monitor the physical assets of the port, the authors suggested data collection from physical access control systems, dedicated monitoring sensors which can create security events, etc. According to the authors, security events from cyber and physical assets should be collected into the hybrid situational awareness framework which can analyse events from both domains. 
    For real-time analysis of these events, two major components were proposed for the situational awareness framework -- Event Correlation Engine (ECE) and Threat Propagation Engine (TPE). According to a widely used definition \cite{jakobson1995}, \emph{event correlation} is a real-time process that involves assigning a new meaning to event groups which occur within predefined time windows. For implementing ECE, the DROOLS engine was used. As the authors stated, the use of ECE allows the detection of complex security incidents which manifest themselves through several events from different domains. The purpose of the TPE was to establish the impact of high-priority security alarms via graph-based analysis, where the nodes of the graph represented cyber and physical assets and the edges of the graph reflected the dependencies between assets.
    The approach by Schauer et al. \cite{9_9032900} was sound and novel considering the time of its publication, integrating cyber and physical situational awareness methods for port infrastructures (including communication with ships). 
    On the other hand, the study did not provide a deeper analysis of the possible cyber attacks and how they could be detected in the maritime infrastructure domain, exemplifying the proposed approach through two attack scenarios.

    The paper 
    \cite{25_10.1145/3290420.3290466} discussed a big data architecture to collect OT data from Unmanned Surface Vehicles (USVs) for various purposes like business management, disaster prevention, anomaly detection, etc. The authors proposed the collection of the following data from USVs -- ship’s performance and navigational data (e.g., speed, position, level of fuel, etc.), data about the environment (e.g., wind, temperature, etc.), data from sensing devices that provided information about the objects around the ship (e.g., data from the radar), and data that was collected from other sources than the USV itself (e.g., industrial data about the USV collected from the manufacturer). For storing the collected data, a cloud-based solution was proposed that was using PostgreSQL database. According to the authors, visualisation capabilities were an important aspect of the big data architecture, and the authors suggested the use of the LinDA toolkit for processing the data in the PostgreSQL database.
    As an example use case for the big data architecture, the paper described an anomaly detection scenario for identifying abnormal combinations of the sailing angle and the USV speed. For anomaly detection purposes, an Isolation Forest-based method was proposed, but the study did not include its performance evaluation.

    Möller et al. \cite{3_8500147} discussed an agent-based intelligent maritime traffic management system to support a coastal Vessel Traffic Service (VTS) with a clear situational picture, traffic and environmental information. For that purpose, the authors proposed to introduce a Multi-Agent System (MAS) that extended to vessels to exchange relevant data (e.g., navigational and safety information) with the central VTS. Following the concept of connected cars, the authors also introduced the concept of connected ships which were continuously in contact with each other. That facilitated the exchange of important navigational information not only through VTS but directly between ships.
    The authors also described how to augment the proposed distributed MAS with a security monitoring functionality. First, the authors detailed the weaknesses of the wireless communication networks used by MAS, and mentioned a mitigation opportunity in the form of anomaly detection for data flows in the network. 
    The study also suggested the use of neural networks for the detection of cyber attacks, but did not report any experiments with the proposed method together with attack detection performance results.

    \textbf{Discussion.} From the four analysed papers, two papers have proposed generic security monitoring architectures, with \cite{1_8602669} discussing a maritime SOC and \cite{9_9032900} a situational awareness system for a port infrastructure which is similar to SOC. The remaining two papers have proposed security monitoring architectures for specific purposes (a big data architecture for monitoring USVs \cite{25_10.1145/3290420.3290466} and a VTS-centric security monitoring architecture \cite{3_8500147}). 
    
    Although the creation of SOCs is a widely studied cyber security topic (see \cite{vielberth} for a recent domain overview), most works in this area lack detailed recommendations and guidelines for implementing a SOC in a scalable and resource-efficient way (for example, recommendations on software solutions to use and guidelines for configuring them) \cite{mases2022}. Furthermore, only a few works have evaluated the SOC architectures and provided relevant performance data (e.g., for resource consumption and event processing rate of system components). Since maritime environments are often resource-constrained, these topics are also relevant in the maritime context and constitute valuable future research directions.

    According to recent studies \cite{alahmadi2022, vielberth}, SOC analysts are often overwhelmed by security alerts of low importance which leads to analyst burnout and alert fatigue. For supporting the human analysts, alert prioritisation has been identified as an important topic for SOC environments \cite{vielberth}, and several ML approaches have been recently proposed for that purpose \cite{vanede2022, guerra2024}. Since alert prioritisation algorithms for reducing human alert fatigue have not been studied in the context of maritime SOC, it can be regarded as one of the open issues for future work.

\subsection{Security monitoring testbeds}\label{testbed-papers}
  
    The short paper 
    \cite{2_7956582} aimed to present a platform to generate data and scenario traces for the evaluation of algorithms for intrusion detection. The introduced solution was based on a ship's two critical subsystems: 1) the propulsion and engine control in which a computer fan simulated the ship's propeller and the fan's actual RPM represented propulsion, 2) the navigation subsystem with the rudder control to simulate the ship's direction changes. Each subsystem had an independent controller which used common industrial communication protocols, like Modbus, DNP3, and S7. The heart of the setup was the central controller built on a Raspberry Pi. The simulated sensor -- that provided the measurement of the engine parts' temperature -- was made of an open-source electronics platform called Arduino. The authors provided an overview of the project, describing its overarching objectives and outcomes, but the overview lacked the required depth 
    for allowing to recreate a similar platform.

    Raimondi et al. \cite{10_9850324} described a testbed for conducting cyber exercises to train maritime SOC teams. The testbed followed the maritime SOC architecture proposed 
    in \cite{1_8602669} (see the previous section for a detailed discussion) and was implemented with Linux containerisation techniques. The testbed employed a ship simulator for simulating the ship at sea and a custom Python script for transmitting data received from the simulator as NMEA messages to other ship-related testbed components. For monitoring the ship network, the authors proposed Suricata IDS which employed Lua scripting for parsing NMEA messages. For the cyber situational awareness console on the ship, the authors used Splunk which also forwarded data to a shore-based central SIEM server (another instance of Splunk). To illustrate the training process of maritime SOC operators, the authors described an example cyber exercise which involved the injection of false NMEA messages into the ship network that interfered with the gyrocompass. The task of the trainees was to detect this attack with the help of Splunk query language.

    In \cite{14_9799429}, a testbed was presented
    for experimenting with the Industrial Control System (ICS) of a warship. To achieve a high degree of realism, the authors utilised the physical devices of real ships as much as possible. The testbed implemented the following four areas of the warship -- propulsion system (engine and propellers), direction (rudders), energy (fuel for the engine and fuel pump), and artillery (76mm main gun and motors for moving the gun turret). For these four areas, a physical implementation contained a bow and rear of the ship with a physical gun turret, rudders, propellers, physical devices of the bridge, etc. According to the authors, the use of virtualised solutions would have decreased the realism of the testbed, since they would have not allowed for the full imitation of the ship’s real-life environment. For low-level monitoring of Modbus RTU fieldbus traffic in the ICS of the ship, the authors proposed the use of Zeek IDS with specialised traffic capturing hardware and Zeek protocol parser for the Fieldbus protocol (in \cite{hotellier2024}, the authors provided a more detailed description of this solution). 
    For regular network monitoring, commercial IDS sensors were suggested by the authors. 

    Basels et al. \cite{43_BaselsTestbed4Radar2024} introduced a testbed to identify navigation radar vulnerabilities and to experiment with relevant security monitoring solutions. The offensive module of the testbed was based on the Radar Attack Tool (RAT) \cite{RadarPWN2022}, which is able to launch a variety of image manipulation attacks against maritime radars (e.g., freezing or scaling the image, removing an object from the image, etc.). The defensive module of the testbed involved Snort3 and IPAL IDS solutions which were configured to detect the attacks against maritime radars by analysing application layer data in network traffic. For Snort3, the authors developed the rules for detecting message injection attacks, whereas IPAL was employed for detecting message modification scenarios. The detection by IPAL relied on two methods outlined in \cite{saillard2024}. 

    Longo et al. \cite{49_LongoTestbed2023} described the MaCySTe testbed designed for maritime cyber security experiments with the vessel IT/OT systems. The authors released detailed installation instructions of the testbed together with the source code in order to facilitate maritime cyber security research in other organisations. The testbed was built on top of Linux containers, making it a highly resource-efficient solution with modest computational requirements. The testbed provided a detailed emulation of the vessel network, navigational devices, propulsion system, and other components together with relevant protocols (e.g., Navico BR24 and Modbus). For evaluating security monitoring functionality, the testbed supported the use of network probes and SIEM. The network probes captured and parsed traffic in the vessel network, forwarding relevant data to SIEM for visualisation and analysis. The authors also provided some illustrative examples on how the MaCySTe testbed can be utilised for cyber attack and attack detection experiments.

    \textbf{Discussion.} From the five analysed papers, two papers have proposed testbeds for experimenting with vessel ICS and SCADA networks \cite{2_7956582, 14_9799429}, one paper has proposed a testbed for maritime radars \cite{43_BaselsTestbed4Radar2024}, one paper has described a more general maritime testbed for a wide range of cyber security experiments \cite{49_LongoTestbed2023}, whereas \cite{10_9850324} focuses on a SOC testbed for training maritime SOC analysts. Since \cite{2_7956582} is a short paper, the discussion of the proposed testbed remains fairly brief, but in the remaining four papers \cite{10_9850324, 14_9799429, 43_BaselsTestbed4Radar2024, 49_LongoTestbed2023} detailed testbed descriptions are provided, with all testbeds being highly realistic environments for conducting maritime cyber security monitoring experiments and trainings.

    However, it should be considered that some of the aforementioned studies represent specific environments (e.g., warship's ICS). Therefore, the creation of highly realistic maritime cyber security testbeds remains an open research area. Also, realistic testbeds allow for generating high-quality datasets for maritime cyber security monitoring experiments. As discussed in Section \ref{taxonomy}, although recent studies have publicly released several maritime-specific datasets, their number remains limited (see Table \ref{security-monitoring-papers-table}), and the creation of such datasets is another important future research direction.

\section{Discussion} \label{discussion}

In order to answer RQ3 (see Section \ref{research_questions}), this section presents the discussion of our main findings about the identified research gaps. The main findings are based on the analysis from the previous section and are presented below.

     \emph{Finding 1: Lack of evaluations on maritime-specific datasets which are publicly available}. Numerous publications have evaluated the proposed security monitoring methods on datasets that are not publicly accessible (see Table \ref{security-monitoring-papers-table}), so the experiments in question are characterised by an inherent lack of reproducibility, thereby precluding the possibility of result verification.
     Although a number of studies have employed publicly available datasets, these publicly available datasets are often not maritime specific but rather generic intrusion detection datasets like NSL-KDD, CICDDoS2019, TON-IoT, etc. (see Table \ref{security-monitoring-papers-table}). Therefore, the validity of these research results remains questionable, because the datasets do not represent the actual maritime systems closely enough. Some notable exceptions include \cite{20_jmse11050928, 42_HossainRadar2024, 45_SinghPopliNet2025, 48_HossainRadar2025} which have employed the publicly available maritime-specific datasets (see Table \ref{security-monitoring-papers-table}).
     
      \emph{Finding 2: Lack of data generation about system failures}. According to recent domain overview papers \cite{ring2019survey, conti2021survey, tantawy2022elements}, research datasets used for developing and testing security monitoring solutions in IT networks \cite{ring2019survey}, ICSs \cite{conti2021survey}, and CPSs \cite{tantawy2022elements} include attack and normal system activities. Thus, the developed security monitoring approaches discriminate attacks from normal activities or identify anomalies originating from malicious actions. We have observed a similar approach in the datasets related to the maritime domain. However, in practice, system failures arising from unintentional causes (e.g., software or hardware errors, environmental condition changes) can be mistakenly detected as anomalies resulting from cyber attacks. Thus, the root causes of the incidents are not usually identified accurately and timely, causing delays and discrepancies in incident handling and recovery operations. As pointed out in \cite{tantawy2022elements}, it is important to include activities representing all failure modes of the target system in the datasets. For example, recent studies in the energy domain aim to discriminate cyber attack related anomalies from non-malicious ones \cite{gupta2022distinguishing, tertytchny2020, wang2021methods}. Therefore, we contemplate that the generation of datasets that include system failures in addition to normal and attack cases would facilitate the research in developing solutions for more granular and informative security monitoring functions.

      \emph{Finding 3: Lack of publicly available experiment code and prototype implementations}. Most studies have not released the implementations of proposed methods and the experiment code (see Table \ref{security-monitoring-papers-table}). This shortcoming is another serious obstacle to achieving reproducibility, and it also complicates the evaluation of the proposed methods on different datasets and in live environments.

     \emph{Finding 4: Lack of proper performance evaluations with appropriate metrics}. Several studies have failed to properly evaluate the performance of the proposed security monitoring methods with proper metrics (see Table \ref{security-monitoring-papers-table}).
     However, in the field of security monitoring and ML, widely acknowledged performance metrics exist such as precision, recall, and F1-score \cite{apruzzese}. The failure to use such widely used metrics makes it difficult to assess what is the attack detection rate and false positive rate of the proposed methods, and how the proposed methods compare to other approaches.

     \emph{Finding 5: Lack of proper computational cost evaluations}. Only some studies (e.g., \cite{15_9478189, 11_9740512, 45_SinghPopliNet2025, 47_AgnewNet2024, 46_LongoRadar2023}) have assessed the computational cost (e.g., CPU time and memory consumption) of the proposed security monitoring methods. However, the computational cost is an important consideration, since vessel networks can usually not accommodate specialised computing platforms for expensive calculations. For example, some recently proposed security monitoring methods are based on DL 
     (see Table \ref{selected-papers-table}),
     which is known to be computationally expensive \cite{apruzzese}, and without the computational cost assessment the applicability of these methods in maritime environments remains questionable. 

     \emph{Finding 6: Lack of detailed assessments of cyber attack impacts on maritime CPSs}. Some papers (e.g., \cite{8_9065023, 16_9728218}) which have proposed methods for detecting cyber attacks against maritime CPS have not conducted cyber attacks in a relevant lab environment. However, the lack of such cyber attack experiments prevents to assess what is the real impact of these attacks on CPS, and what are the best avenues for detecting these attacks. Furthermore, the assessment of cyber attack impacts and best attack detection approaches requires a lab environment which mimics CPS as closely as possible. However, apart from a few recent studies \cite{14_9799429, 49_LongoTestbed2023}, the methods for building highly realistic maritime CPS labs and testbeds have not received enough attention in the relevant literature.

     \emph{Finding 7: Lack of detailed evaluation of the proposed security monitoring system architectures}. Security incident handling processes usually require collecting, aggregating and correlating security events from various sources to do more relevant prioritisation, accurate incident categorisation and impact assessment. 
     Although the idea of using maritime SOC and aggregating diverse security data from many sources has been proposed in several past studies \cite{1_8602669, 9_9032900, 3_8500147, 25_10.1145/3290420.3290466}, existing works have not published detailed evaluation data for the proposed architectures. However, some similar studies that are not maritime specific have included such evaluations (e.g., see \cite{mases2022}), and works from maritime domain should follow the same approach.

     \emph{Finding 8: Security monitoring methods for some maritime systems have been understudied}. As discussed in Section \ref{findings}, security monitoring methods for distributed maritime networks have been largely focusing on IoT networks, while other network types have received less attention. Also, most network monitoring research has utilised supervised ML methods, and studying the feasibility of other methods is a valuable future research direction. For example, semi-supervised ML is known to greatly reduce the model development cost of supervised ML \cite{apruzzese2}, but we found only one relevant work in the analysed literature 
     \cite{42_HossainRadar2024}.
     Also, process-aware maritime CPS monitoring which correlates indicators from the physical and cyber space remains an open research topic. As for security monitoring methods for navigational systems, the use of supervised and semi-supervised ML, rule mining, and inductive logic programming for rule learning are other possible future research directions.
     
     \emph{Finding 9: Advanced ML issues have been understudied in the maritime context}. The shortage of domain-specific datasets in the maritime field is a major obstacle to understanding the performance of ML methods in solving maritime security monitoring problems. However, once datasets are created, various advanced problems related to the application of ML methods can be addressed. Handling the concept drifts that may occur in the feature space, learning from small labelled datasets and efficient utilisation of experts for labelling are some research directions that can be tackled in this domain. The application of large language models for a better understanding of the detection rules and monitoring results \cite{juttner2023chatids}, and receiving suggestions from these models about the course of actions can be studied in the maritime context. Some of the security monitoring tasks can be deployed into the resource-constraint devices in maritime systems for making real-time decisions. Thus, another potential research dimension would address running and optimising learning models on such devices (e.g., TinyML applications as recommended in a recent study \cite{morariu2024}).

     \emph{Finding 10: Human aspects of security monitoring systems have been understudied in the maritime context}. Although the ultimate goal of cyber security monitoring would be the full automatisation of attack detection and response mechanisms, it is highly expected that human experts will take a critical part in incident handling processes in any case. Thus, interdisciplinary and holistic approaches involving human and technical aspects of the problem domain are highly needed. Training and awareness programs for security analysts and maritime operators addressing design issues for the minimisation of human errors (e.g., optimising user interfaces of monitoring tools), enhancing communication between incident handling teams or selecting the right course of action under stressed conditions are some sample topics that require further research in this domain. 
     Another potential research  topic is the use of ML tools for alert prioritisation that helps to prevent alert fatigue among SOC analysts
     (a recent study among maritime cyber security experts has confirmed the importance of this topic \cite{nganga2024sensors}).
     Apart from a few works which have discussed the expected competence of maritime SOC analysts and their training process \cite{10_9850324}, 
     and the design of training events for building maritime cyber security skills \cite{harish2024},
     existing studies have generally not touched on human aspects of maritime security monitoring.

\section{Conclusion and future work} \label{conclusion}
    In our research, we conducted a systematic literature review in which we surveyed four digital libraries. After a careful selection process described in Section \ref{methodology}, we identified 32 papers in the field of cyber security monitoring in maritime. 
    In Section \ref{findings}, we provided a comprehensive overview of these publications, highlighting their strengths and shortcomings. We shared the bibliometric analysis of the papers in Section \ref{bibliometry}, and we discussed the identified research gaps in Section \ref{discussion}.
    Our developed taxonomy (outlined in Section \ref{taxonomy}) provides a structured framework for analysing cyber security monitoring-related literature, allowing insights into the strengths and limitations of existing publications. 
     
     As we discussed in Section \ref{discussion}, we identified several shortcomings and limitations of the currently available literature. The present study yields the following findings. 
     First, we found several dataset related issues (Findings 1-2). Maritime-specific publicly available datasets should be used for experiments to ensure the reproducibility and the validity of research results for maritime environments. Also, the datasets should include system failure data not associated with cyber attacks, since that would allow us to evaluate how well the security monitoring methods can distinguish the effects of cyber attacks from other system failures.

     Second, we identified several evaluation related issues (Findings 3-7). To improve the quality of evaluations, appropriate metrics should be used for assessing the performance of algorithms. Also, it is important to assess the computational cost of the proposed methods. Evaluations should be realistic, involving real cyber attacks that are conducted in a lab environment with a high degree of realism. For the sake of reproducibility, it is essential to share the experiment code and prototype implementations. Similarly to security monitoring methods, detailed evaluations are needed for security monitoring system architectures.

     Third, we identified several previously unexplored research areas of maritime security monitoring (Findings 8-10). 
     For example, the monitoring methods for maritime IoT networks and vessel propulsion systems have been studied more in the current literature, whereas other types of maritime distributed networks and maritime CPSs are potential targets of future research. Similarly,
     process-aware model-based methodologies, emphasising the correlation between physical and cyberspace indicators for enhanced system malfunction and cyber attack distinction, could be leveraged in the maritime domain. Furthermore, several advanced ML issues like concept drift and learning models for resource-constrained environments deserve more attention in the field of maritime security monitoring. While the training of personnel constitutes a fundamental aspect of cyber security monitoring, it is noteworthy that only few works specifically addressing this subject matter could be identified in the relevant literature.

    As for future work, we plan to study the methods of creating a realistic maritime cyber security monitoring lab, using our experience in the field \cite{visky2022}. Also, we plan to use this lab environment for the creation and release of maritime-specific datasets to facilitate security monitoring research in the maritime domain. Finally, our plans include studying advanced ML-based security monitoring methods on these datasets.

\section*{Acknowledgment}
This research was funded by the EU Horizon2020 project MariCybERA (agreement No 952360).


\end{document}